\documentstyle[12pt,epsfig]{article}

\newlength{\dinwidth}
\newlength{\dinmargin}
\def\lapproxeq{\lower .7ex\hbox{$\;\stackrel{\textstyle
<}{\sim}\;$}}
\def\gapproxeq{\lower .7ex\hbox{$\;\stackrel{\textstyle
>}{\sim}\;$}}

\def\be{\begin{equation}}
\def\ee{\end{equation}}
\def\bea{\begin{eqnarray}}
\def\eea{\end{eqnarray}}

\def\funp{{I\!\!P}}

\begin{document}
\titlepage
\begin{flushright}
IPPP/01/53 \\
DCPT/01/106 \\
20 February 2002 \\
\end{flushright}

\vspace*{2cm}

\begin{center}
{\Large \bf Prospects for New Physics observations in diffractive}

\vspace{0.4cm}

{\Large \bf processes at the LHC and Tevatron} \\

\vspace*{1cm}
V.A. Khoze$^a$, A.D. Martin$^a$,  and M.G. Ryskin$^{a,b}$ \\

\vspace*{0.5cm}
$^a$ Department of Physics and Institute for Particle Physics Phenomenology, University of
Durham, Durham, DH1 3LE \\
$^b$ Petersburg Nuclear Physics Institute, Gatchina, St.~Petersburg, 188300, Russia
\end{center}

\vspace*{1cm}

\begin{abstract}
We study the double-diffractive production of various heavy
systems (e.g.\ Higgs, dijet, $t\bar{t}$ and SUSY particles) at LHC
and Tevatron collider energies.  In each case we compute the
probability that the rapidity gaps, which occur on either side of
the produced system, survive the effects of soft rescattering and
QCD bremsstrahlung effects.  We calculate both the luminosity for
different production mechanisms, and a wide variety of subprocess
cross sections.  The results allow numerical predictions to be
readily made for the cross sections of all these processes at the
LHC and the Tevatron collider.  For example, we predict that the
cross section for the {\it exclusive} double-diffractive
production of a 120~GeV Higgs boson at the LHC is about 3~fb, and
that the QCD background in the $b\bar{b}$ decay mode is about 4
times smaller than the Higgs signal if the experimental
missing-mass resolution is 1~GeV.  For completeness we also
discuss production via $\gamma\gamma$ or $WW$ fusion.
\end{abstract}

\newpage

\section{Introduction}

Double-diffractive processes of the type
\begin{equation}
\label{eq:a1}
 p p \; \rightarrow \; X + M + Y,
\end{equation}
can significantly extend the physics programme at high energy
proton colliders.  Here $M$ represents a system of invariant mass
$M$, and the $+$ signs denote the presence of rapidity gaps which
separate the system $M$ from the products $X$ and $Y$ of proton
diffractive dissociation.  Such processes allow both novel studies
of QCD at very high energies and searches for New Physics. From a
theoretical point of view, hadronic processes containing rapidity
gaps play a crucial role in determining the asymptotic behaviour
of the cross section at high energies.  From an experimental
viewpoint, the presence of rapidity gaps provides a clean
environment for the identification of a signal.  In such events we
produce a colour-singlet state $M$ which is practically free from
soft secondary particles.

Double-diffractive {\it exclusive} processes of the type
\begin{equation}
\label{eq:b1}
 pp \; \rightarrow \; p \: + \: M \: + \: p,
\end{equation}
where the protons remain intact, are even better.  They allow the
reconstruction of the ``missing'' mass $M$ with good resolution,
and so provide an ideal way to search for new resonances and
threshold behaviour phenomena.  Moreover, in exclusive processes
with forward protons, as shown in Fig.~1(a), the incoming $gg$
state satisfies special selection rules, namely it has $J_z = 0$,
and positive $C$ and $P$ parity.  Hence only a subset of resonant
states $M$ can be produced, in particular $0^{++}$ (but not, for
example, $1^{++}$).  Furthermore, the selection rules control the
threshold behaviour of the production of a pair of heavy
particles.  For example, the Born cross section for the production
of a fermion-antifermion pair is proportional to $\beta^3$, where
$\beta$ is the fermion velocity, while for scalar particles the
threshold behaviour is just $\beta$.  Thus, for instance, in this
way we can distinguish between scalar quark, $\tilde{q}$, and
gluino, $\tilde{g}$, pair production.  Also, the selection rules
are of crucial importance in suppressing the $b\bar{b}$ QCD
background when searching for the $H \rightarrow b\bar{b}$ signal
\cite{BBB,KMRmm}.

We may write the cross sections for processes (\ref{eq:a1}) and
(\ref{eq:b1}) in the factorized form
\begin{equation}
\label{eq:a2}
 \sigma \; = \; {\cal L} (M^{2}, y) \: \hat{\sigma} (M^{2}),
\end{equation}
where $\hat{\sigma}$ is the cross section for the hard subprocess
which produces the system of mass $M$, and $\cal L$ is the
effective luminosity for production at rapidity $y$. The
luminosity $\cal L$ may refer to any colourless $t$-channel states
which transfer energy across the rapidity gaps without radiating
secondary particles.  For example, the colourless exchange may be
a two-gluon `hard Pomeron' state, or a phenomenological `soft'
Pomeron, or even a $\gamma$ or $W$ boson.

The aim of the present paper is to provide a comprehensive way to
estimate the numerical size of the cross sections for all these
double-diffractive mechanisms for producing a wide range of heavy
systems $M$ (for example, Higgs, dijet, $t\bar{t}$ and SUSY
particles).  First, in Section~2 we determine the luminosity for
the different double-diffractive production mechanisms.  We
present plots which readily give the luminosity as a function of
the mass $M$ and rapidity $y$ of the heavy system, for $pp
(\bar{p}p)$ collider energies $\sqrt{s} = 2, 8$ and 14~TeV.  Then
in Section~3 we give the formulae for, and numerical estimates of,
the hard subprocess cross section $\hat{\sigma}$ for the different
heavy systems $M$.  Armed with this numerical information of
${\cal L}$ and $\hat{\sigma}$, we can immediately estimate the
observable cross sections $\sigma$ of a wide variety of different
double-diffractive processes at the LHC and the Tevatron, see
(\ref{eq:a2}).

We consider the following double-diffractive mechanisms.
Production of the heavy system
\begin{itemize}
\item[(a)] via the exclusive process $p p \rightarrow p + M + p$,
shown in (\ref{eq:b1}),
\item[(b)] via the inclusive process $pp \rightarrow X + M + Y$, shown in (\ref{eq:a1}),
\item[(c)] via the inelastic collision of two Pomerons.
\end{itemize}
The three different mechanisms are shown in Fig.~1. Since we are
considering the production of a massive system $M$ at high
collider energy, the luminosity of the first two mechanisms may be
calculated from perturbative QCD, using the known parton
distributions of the proton.  Although the cross section
$\hat{\sigma}$ for the hard subprocess is factored off, recall
that there may exist specific selection rules for the different
configurations.  In particular in the case of exclusive production
only the projection with $J_{z} = 0$ and $P = +1$ contributes to
$\hat{\sigma}$.

\begin{figure}[!h]
\begin{center}
\epsfig{figure=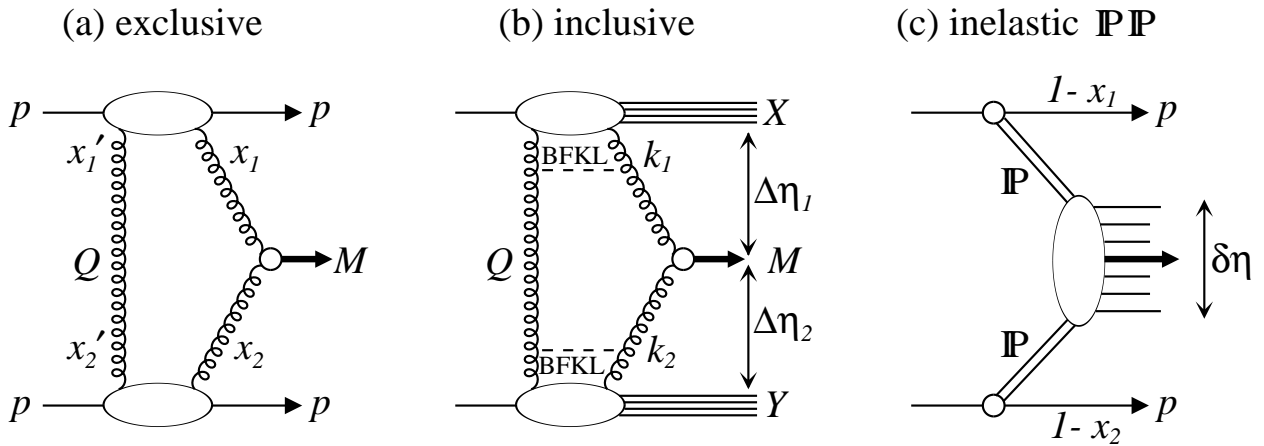,height=2.5in}
 \caption{Different mechanisms for the double-diffractive production of a system of
mass $M$ in high energy proton-proton collisions.}
\label{martin-Fig1}
\end{center}
\end{figure}

We need to be careful in interpreting the cross section
$\hat{\sigma}$ for the subprocess producing the massive system,
$gg \rightarrow M$.  It corresponds to {\it colour singlet}
production.  That is the subprocess amplitude is averaged over the
colour indices of the incoming gluons
\begin{equation}
\label{eq:b2}
 {\cal M} \; = \; \frac{1}{N_C^2 - 1} \: \sum_{a, b} \: {\cal
 M}_{ab} \: \delta_{ab},
\end{equation}
where $N_C = 3$ and ${\cal M}_{ab}$ is the amplitude for the
fusion subprocess $g^{a} g^{b} \rightarrow M$.  All other colour
factors are included in the luminosity.  Moreover, in the case of
exclusive double-diffractive production, the {\it amplitudes}
(rather than the cross sections) are averaged over the two {\it
transverse} polarisations of the incoming gluons
\begin{equation}
\label{eq:c2}
 {\cal M} \; = \; \frac{1}{2} \: \sum_{\varepsilon_1,
 \varepsilon_2}\: {\cal M}^{\varepsilon_1, \varepsilon_2} \: \delta_{\varepsilon_1 \varepsilon_2}.
\end{equation}
As a consequence, at leading order, we have only $J_z = 0$
production.  On the contrary, for inclusive production, the
subprocess {\it cross section} is averaged over {\it all}
polarisation states of the incoming gluons, as usual.  Finally
note that for production via soft Pomeron-Pomeron fusion, the
cross section $\hat{\sigma}$ includes the convolution with the
parton distributions of the Pomeron.

\section{Luminosities for double-diffractive processes}

In this section we collect together the formulae necessary to
compute the luminosity functions for the processes shown in
Fig.~1. We write the effective\footnote{The Pomeron should be
regarded as including all multi-Pomeron effects.} luminosities,
for producing a system of mass $M$ and rapidity $y$, in the form
\begin{equation}
\label{eq:a3}
 M^2 \: \frac{\partial {\cal L}^{(i)}}{\partial y \:
\partial M^2} \; = \; \hat{S}^{2 (i)} \: L^{(i)},
\end{equation}
with $i = a,b,c$ corresponding to the processes shown in Fig.~1.
Here we have integrated over the transverse momenta of the
outgoing protons or outgoing proton dissociated systems, according
to whether we are considering exclusive or inclusive double
diffractive production.  An important ingredient in the
calculation of the luminosity is the inclusion of the survival
probability of the rapidity gaps to, first, soft rescattering of
the interacting protons and, second, to QCD radiation.  The
latter, which results in a Sudakov-like suppression, is included
in the expressions for $L^{(i)}$ below.  The soft rescattering
effects are symbolically denoted by a factor $\hat{S}^2$ in
(\ref{eq:a3}).  In practice, we calculate the effects using a
two-channel eikonal.  As a consequence (\ref{eq:a3}) does not have
a factorized form, and $\hat{S}^2$ should be viewed as the soft
survival probability appropriately averaged over the channels
\cite{KMRsoft,KKMR}.  The value of the survival factor $\hat{S}^2$
depends on the particular double-diffractive production mechanism,
and may be a function of $y$ and $M^2$.

\subsection{Exclusive double-diffractive production}

For the exclusive process shown in Fig.~1(a) we have, to single
$\log$ accuracy, \cite{KMR}
\begin{equation}
\label{eq:a4}
 L^{\rm excl} \; = \; \left ( \frac{\pi}{(N_C^2 - 1) b} \int \frac{dQ_t^2}{Q_t^4}
 \: f_g (x_1, x_1^\prime, Q_t^2, \mu^2) \: f_g (x_2, x_2^\prime,
 Q_t^2, \mu^2) \right )^2,
\end{equation}
where $b$ is the $t$-slope corresponding to the momentum transfer
distributions of the colliding protons
\begin{equation}
\label{eq:a5}
 \frac{d^2 \sigma}{dt_1 dt_2} \; \propto \; e^{b (t_1 + t_2)},
\end{equation}
where we take\footnote{If we were to adopt a Regge interpretation,
then the `Pomeron' would be represented in (\ref{eq:a4}) by an
unintegrated gluon distribution $f_g$ which, for a large hard
scale $\mu$, is described by DGLAP evolution.  As a consequence of
the strong $k_t$ ordering, the position of the gluons in impact
parameter space is frozen, and hence there is no shrinkage of the
diffraction cone.  This would mean that the corresponding Pomeron
trajectory would have zero slope, $\alpha^\prime = 0$.  Therefore
we choose a constant $t$-slope, $b$, which characterises the $t$
dependence of the Pomeron-proton vertex.  The value $b = 4~{\rm
GeV}^{-2}$ is taken from the fit to the soft hadronic data of
Ref.~\cite{KMRsoft}.  This value is consistent with that observed
in $J/\psi$ diffractive production at HERA \cite{JPSI}.} $b =
4~{\rm GeV}^{-2}$. The quantities $f_g (x, x^\prime, Q_t^2,
\mu^2)$ are the generalised (skewed) unintegrated gluon densities
of the protons. The skewed effect arises because the screening
gluon ($Q_t$) carries a much smaller momentum fraction $x^\prime
\ll x$. For small $|x - x^\prime|$ the skewed unintegrated density
can be calculated from the conventional integrated gluon $g (x,
Q_t^2)$ \cite{MR}. However the full prescription is rather
complicated. For this reason it is often convenient to use the
simplified form \cite{KMR}
\begin{equation}
\label{eq:a6}
 f_g (x, x^\prime, Q_t^2, \mu^2) \; = \; R_g \: \frac{\partial}{\partial \ln
 Q_t^2}\left [ \sqrt{T (Q_t, \mu)} \: xg (x, Q_t^2) \right ],
\end{equation}
which holds to 10--20\% accuracy.  The factor $R_g$ accounts for
the single $\log Q^2$ skewed effect \cite{SGMR}.  It is found to
be about 1.2 at the LHC energy, and 1.4 at the Tevatron energy.
The Sudakov factor $T (Q_t, \mu)$ is the survival probability that
a gluon with transverse momentum $Q_t$ remains untouched in the
evolution up to the hard scale $\mu = M/2$
\begin{equation}
\label{eq:a7}
 T (Q_t, \mu) \; = \; \exp \left ( - \int_{Q_t^2}^{\mu^2} \:
 \frac{\alpha_S (k_t^2)}{2 \pi} \: \frac{dk_t^2}{k_t^2} \:
 \int_0^{1 - \Delta} \: \left [zP_{gg} (z) \: + \: \sum_q \:
 P_{qg} (z) \right ] \: dz \right ),
\end{equation}
with $\Delta = k_t/(\mu + k_t)$.  The square root arises in
(\ref{eq:a6}) because the survival probability is only relevant to
the hard gluon.  It is the presence of this Sudakov factor which
makes the integration in (\ref{eq:a4}) infrared stable, and
perturbative QCD applicable.

\begin{figure}[!ht]
\begin{center}
\epsfig{figure=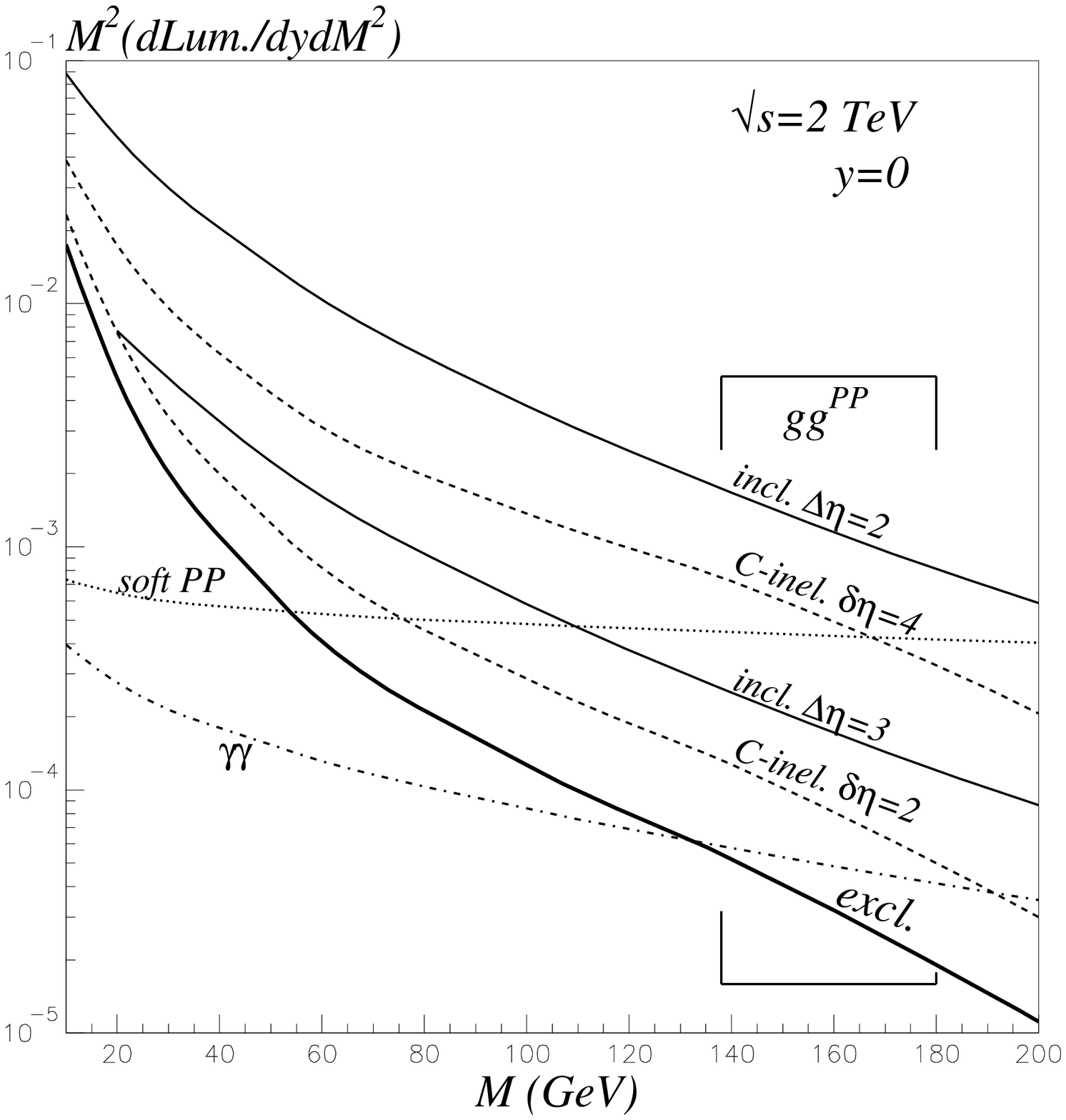,height=3.3in}
\epsfig{figure=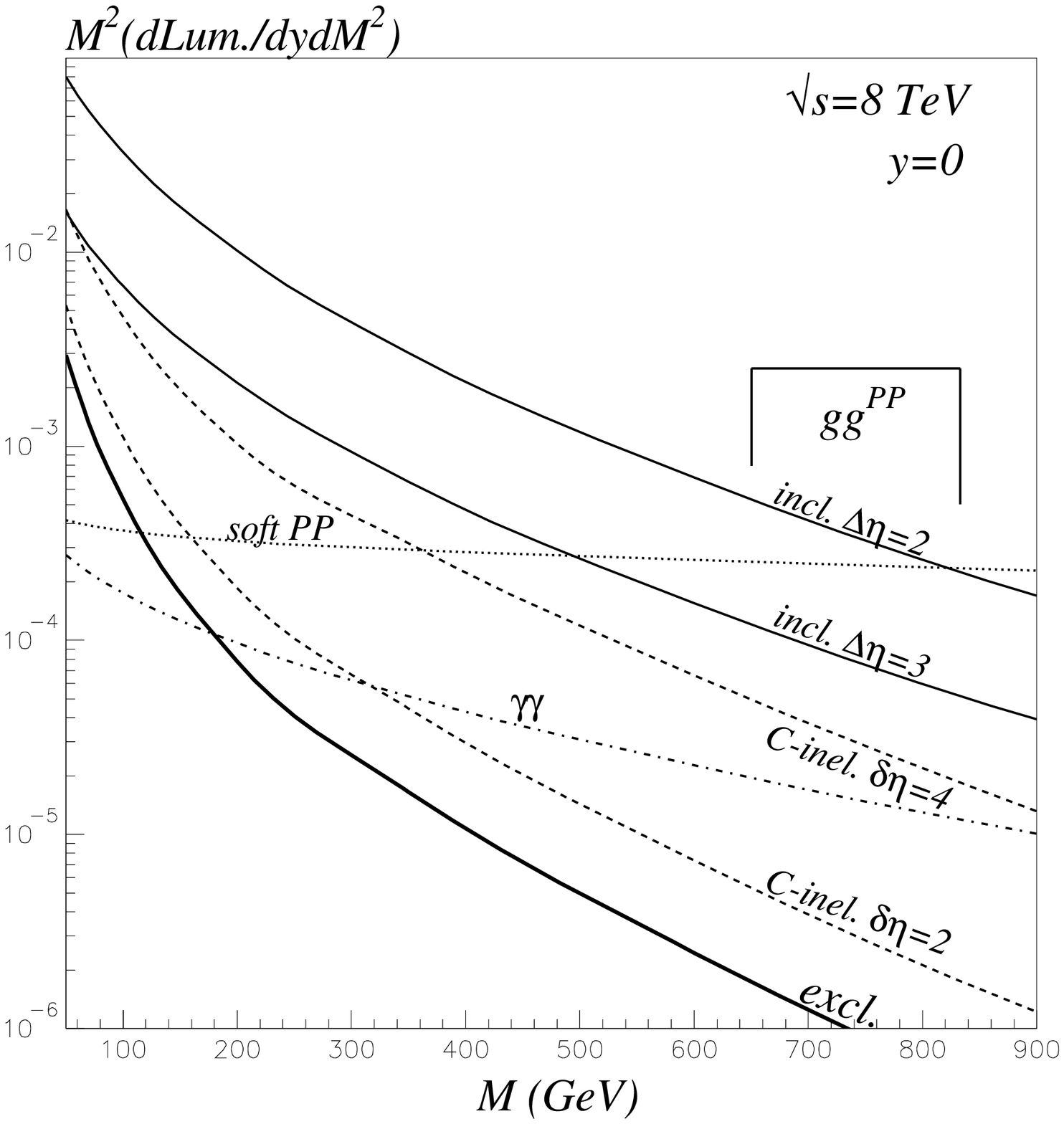,height=3.3in}
\epsfig{figure=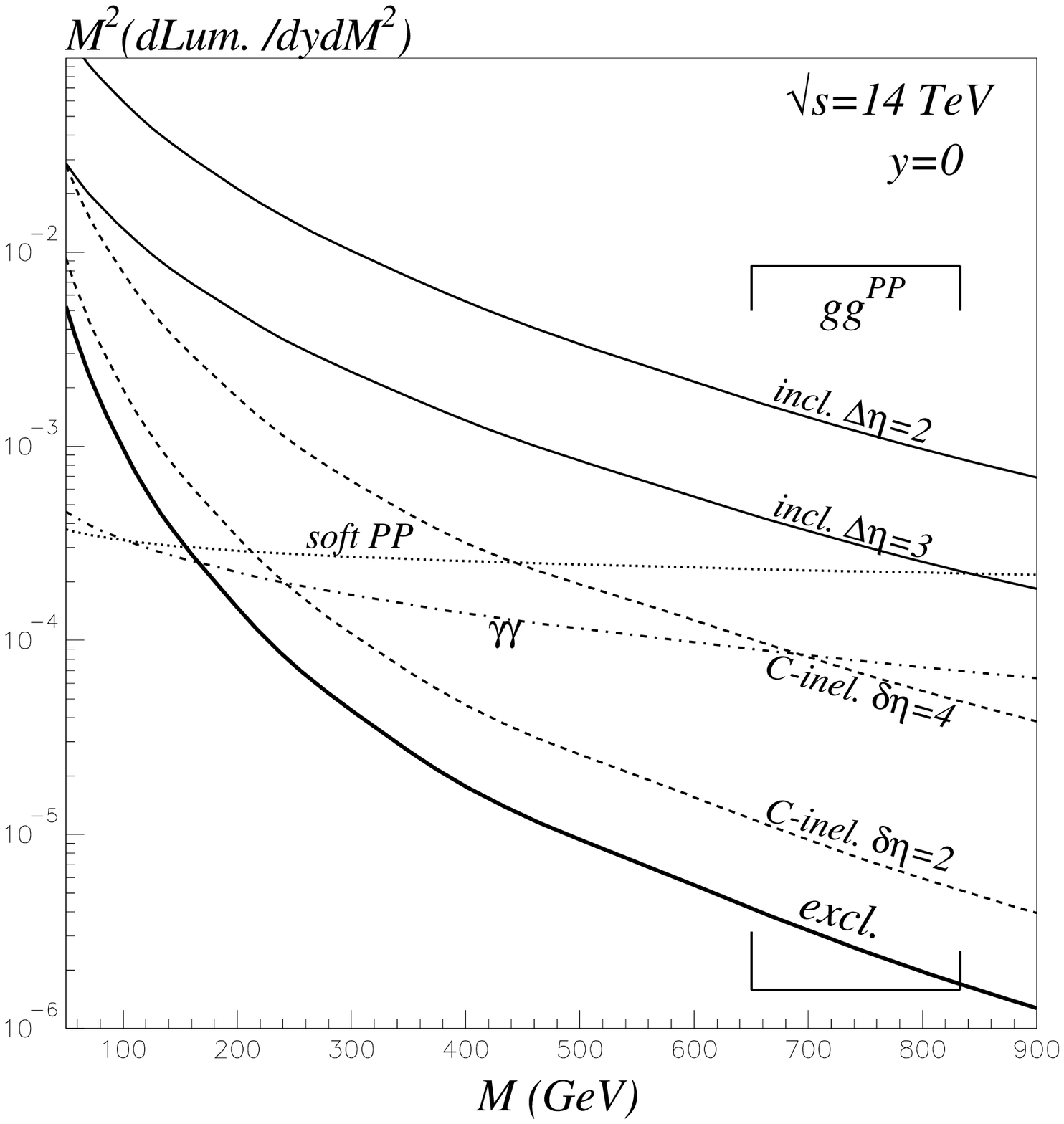,height=3.3in}
  \caption{The luminosity $M^2 \partial {\cal L}/\partial y \partial M^2$
 versus $M$, for the double-diffractive production of a heavy system of
 mass $M$ with rapidity $y = 0$.  The three plots are for $pp$ (or
 $p\bar{p}$) collider energies of $\sqrt{s} = 2, 8$ and 14~TeV.
 Various production mechanisms are studied: the curves marked {\it excl.},
 {\it incl.}, {\it C-inel} and {\it soft} $\funp\funp$ correspond,
 respectively, to production by the exclusive process
 $pp \rightarrow p + M + p$ of Section~2.1, to production by the inclusive
 process $pp \rightarrow X + M + Y$ of Section~2.2, and to production by the
 processes shown in Figs.~6(a) and 6(b) as described in Section~2.3.  The
 $\gamma\gamma$ luminosity is obtained as described in Section~2.4.}
 \label{Fig2}
\end{center}
\end{figure}

\begin{figure}[!h]
\begin{center}
\epsfig{figure=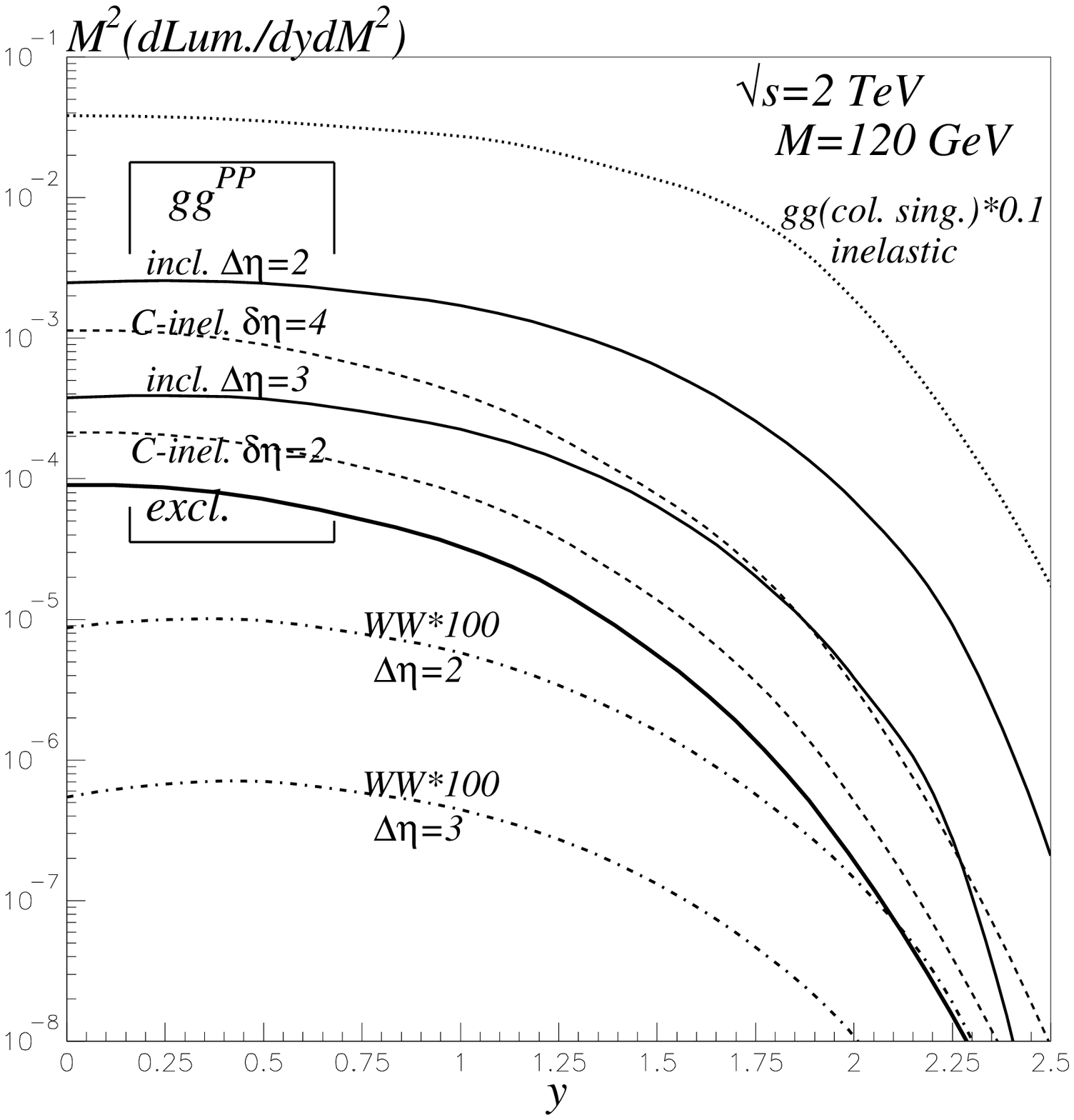,height=3.3in}
\epsfig{figure=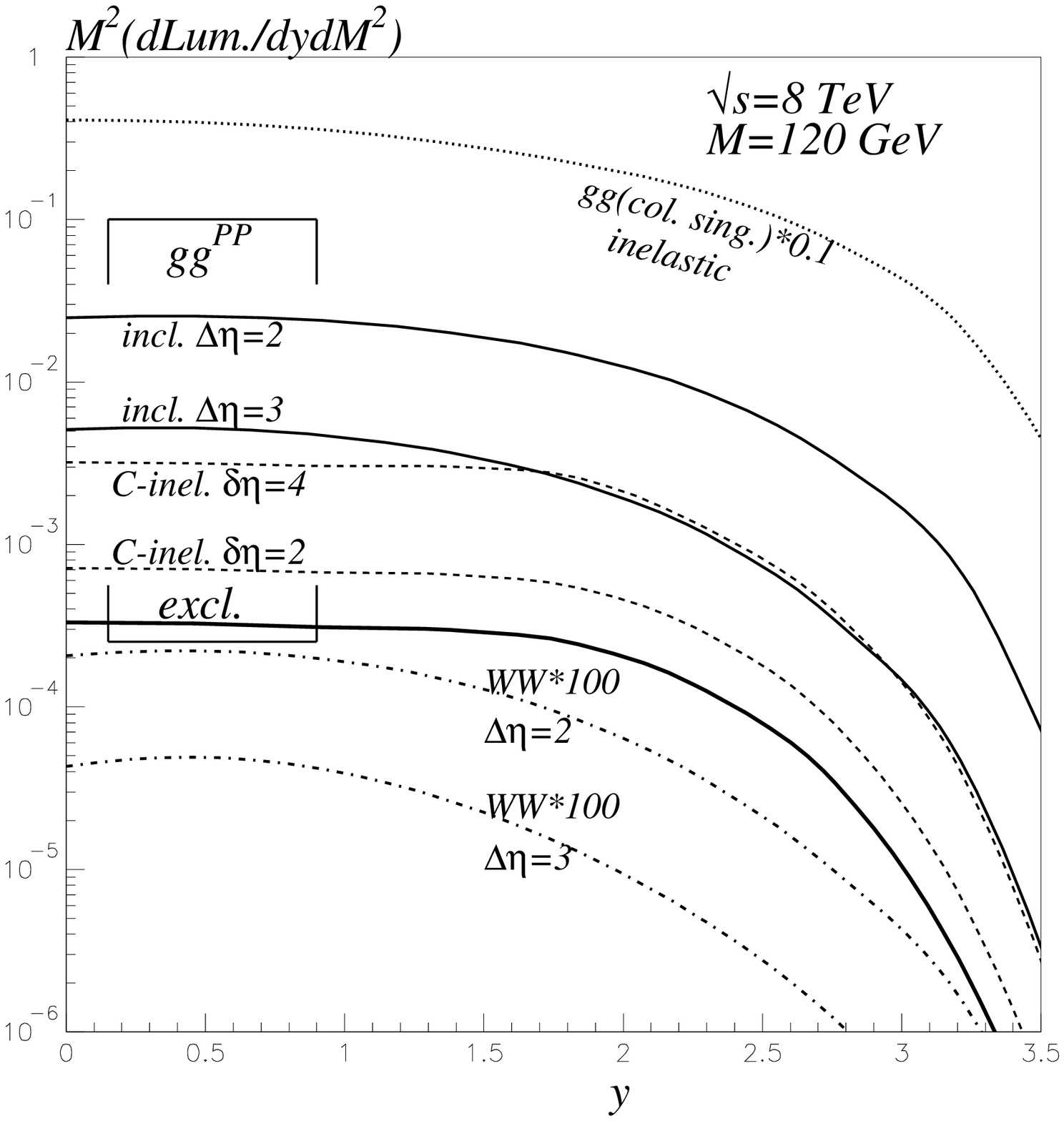,height=3.3in}
\epsfig{figure=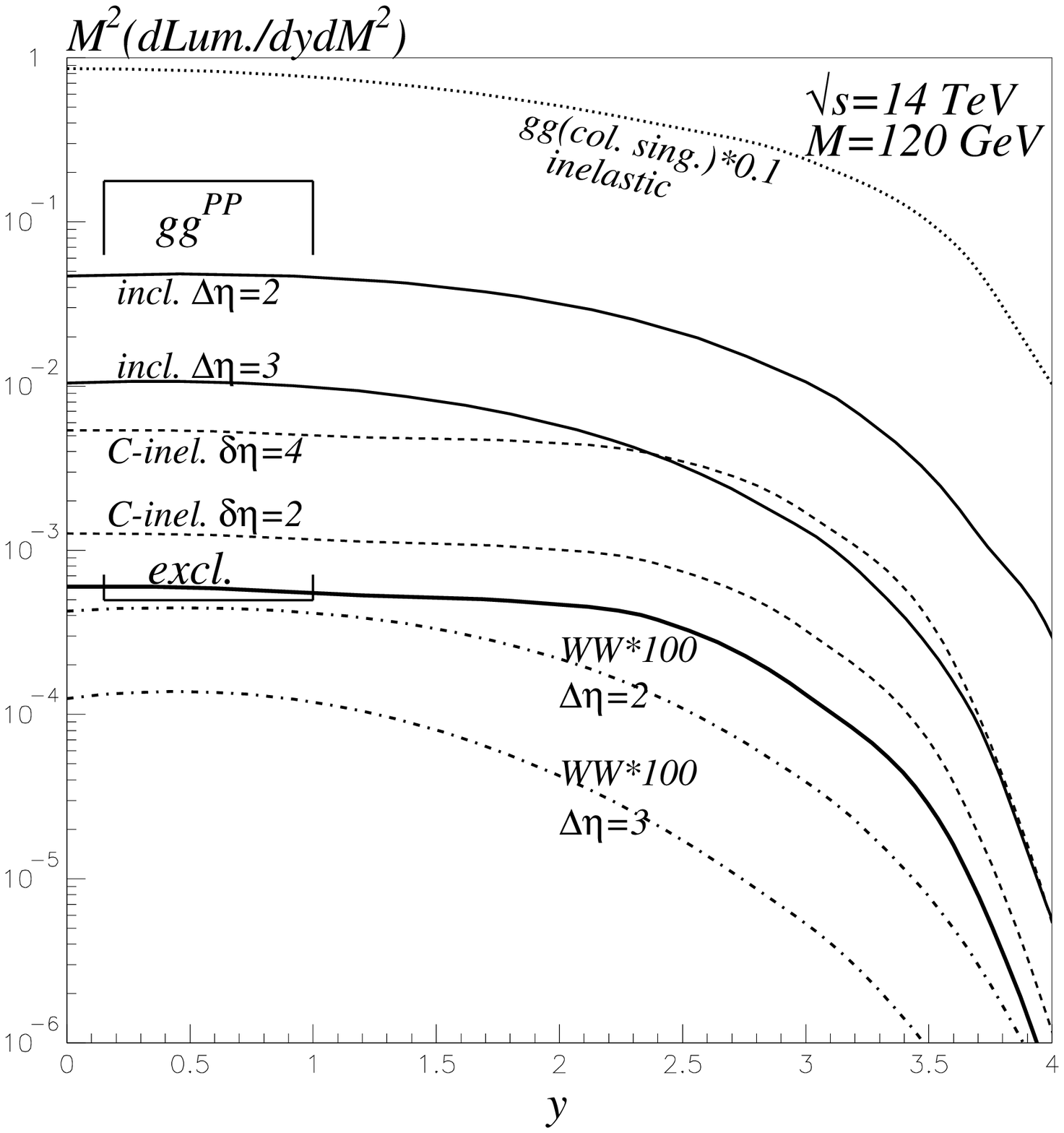,height=3.3in}
 \caption{The luminosity $M^2 \partial {\cal L}/\partial y \partial
 M^2$ versus $y$ for the double-diffractive production by various
 mechanisms of a heavy system of mass $M = 120$~GeV at $\sqrt{s} = 2, 8$ and 14~TeV.
 The notation for the  curves is as in Figs.~2 and 6.  The upper curve in each plot shows the
 inelastic luminosity $(\Delta \eta = 0)$ assuming that the fusing $gg$
 pair are in a colour singlet state.}
 \label{Fig3}
\end{center}
\end{figure}

\begin{figure}[!h]
\begin{center}
\epsfig{figure=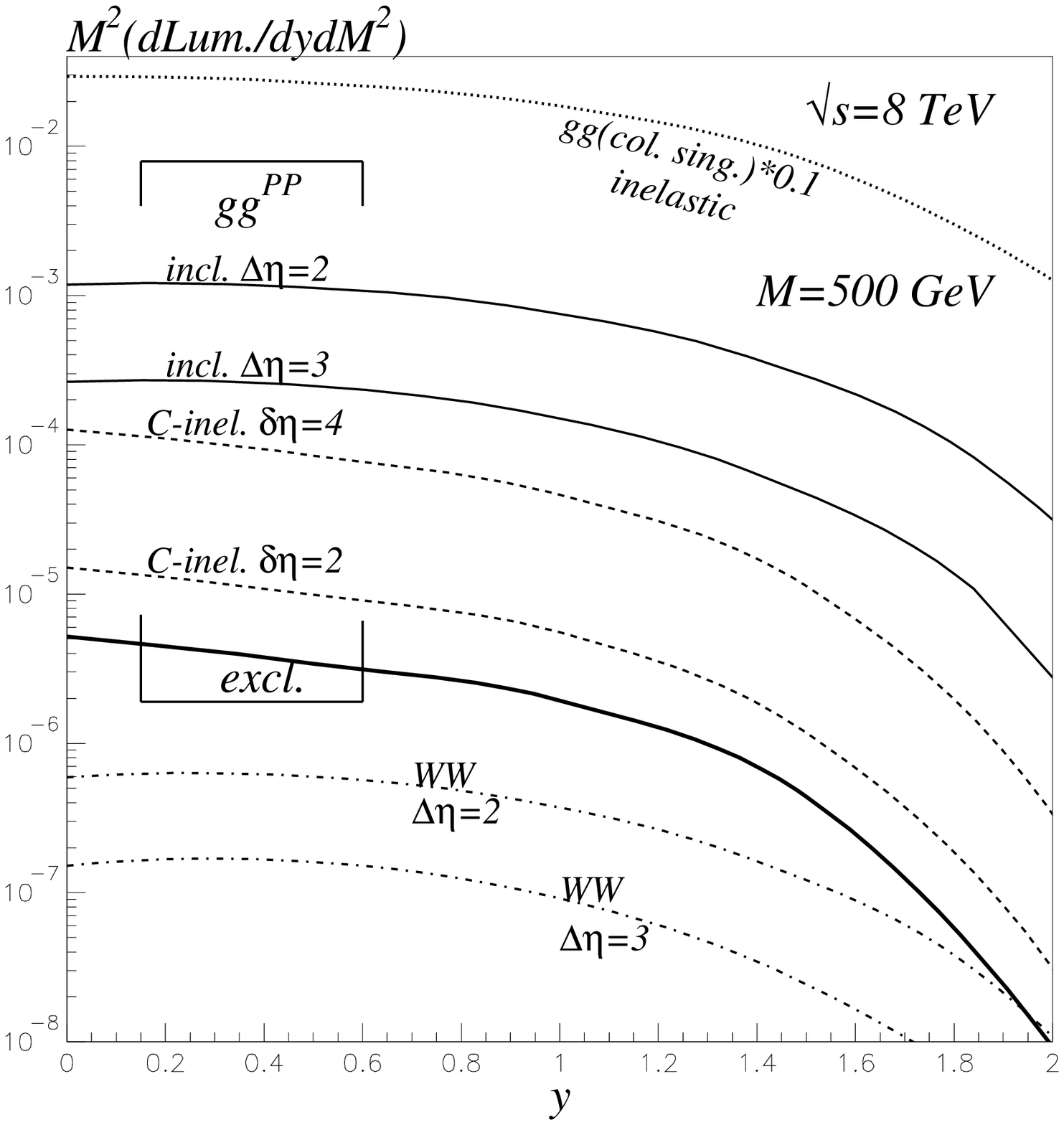,height=3.3in}
\epsfig{figure=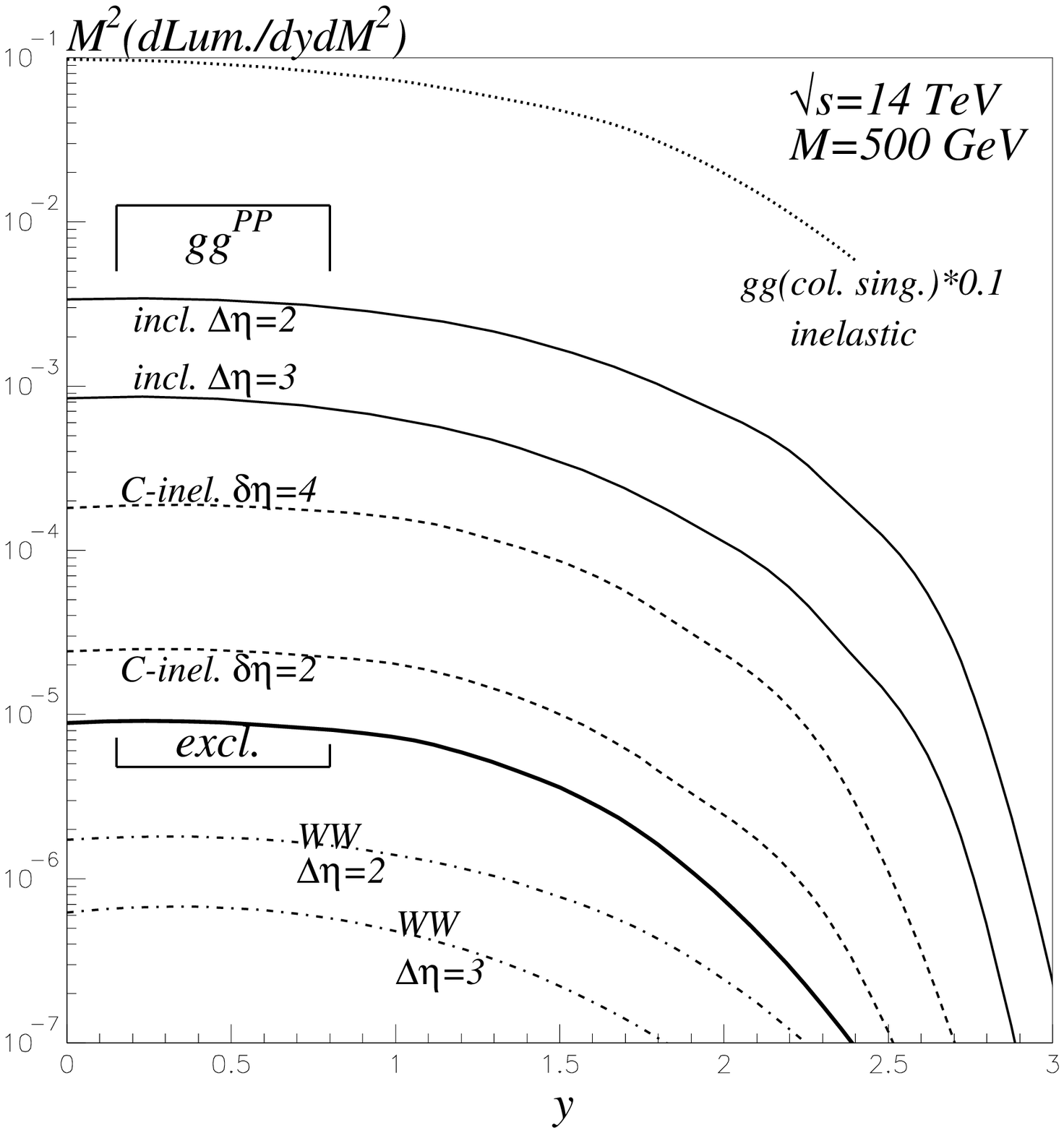,height=3.3in}
 \caption{As for Figure~3, but for the production of a system of mass $M = 500$~GeV at
 $pp$ collider energies of $\sqrt{s} = 8$ and 14~TeV.}
 \label{Fig4}
\end{center}
\end{figure}

The luminosities $M^2 d{\cal L}/dy dM^2$ calculated from
(\ref{eq:a4}) and (\ref{eq:a3}) for the exclusive
double-diffractive production of a system of invariant mass $M$
and rapidity $y$ are shown by the solid continuous curves, denoted
{\it excl}, in Figs.~2, 3 and 4.  Fig.~2 shows the dependence of
the luminosity on $M$ at $y = 0$, and Figs.~3 and 4 show the
dependence on $y$ for produced masses of $M = 120$ and $M =
500$~GeV respectively, at various $pp$ (or $p\bar{p}$) collider
energies. We use the MRST99 partons \cite{MRST} to calculate the
unintegrated gluon distributions (\ref{eq:a6}), and we calculate
the `soft' survival factor $\hat{S}^2$ using the formalism of
Ref.~\cite{KMRsoft}.  We find $\hat{S}^2 = 0.045$, 0.026 and 0.020
for collider energies $\sqrt{s} = 2, 8$ and 14~TeV respectively.

The luminosity is presented as the number of effective gluon-gluon
collisions per $pp$ interaction.  In Figs.~2 and 3 we denote it as
$gg^{PP}$ to indicate that the hard gluons, which interact to form
the system $M$, originate within overall colourless (hard Pomeron)
$t$-channel exchanges, see Fig.~1(a).  This is precisely the
quantity which must be multiplied by the cross section
$\hat{\sigma}$ of the $J_z = 0$, colour-singlet hard subprocess
$gg \rightarrow M$, to form the double-diffractive cross section
$\sigma$ of (\ref{eq:a2}).

As can be seen from Fig.~2 the luminosity decreases with
increasing $M$, but grows with the collider energy $\sqrt{s}$. The
reason is that there is an increasing number of gluons as $x$
becomes smaller, and because for larger $M$ the double-logarithmic
Sudakov suppression (\ref{eq:a7}) is stronger due to the higher
scale $\mu = M/2$.

\subsection{Inclusive double-diffractive production}

\begin{figure}[!h]
\begin{center}
\epsfig{figure=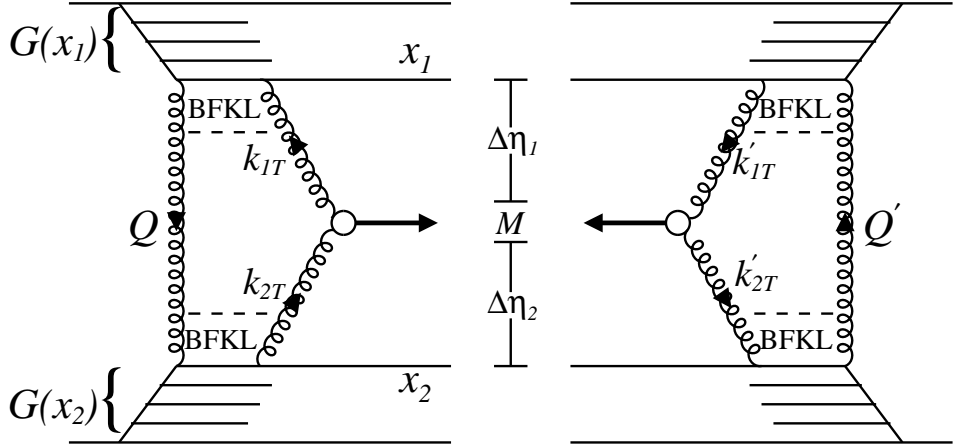,height=2.5in}
 \caption{The amplitude of Fig.~1(b) multiplied by its complex conjugate,
 which gives the cross section for the inclusive double-diffractive production
 of a system $M$.  The effective parton densities, $G (x_i)$ of
 (\ref{eq:a10}), are to be evaluated at scales $k_{it}^2$.
 \label{Fig5}}
\end{center}
\end{figure}

Exclusive production has by far the cleanest signal.
Unfortunately, the luminosity, and hence the predicted event rate,
are small.  For this reason we consider {\it inclusive} double
diffractive production, process (\ref{eq:a1}), with less
restrictive kinematics due to the dissociation of the incoming
protons, see Fig.~1(b).  The cross section can be expressed as
exclusive production at the parton-parton level
\begin{equation}
\label{eq:b7}
 a_1 a_2 \; \rightarrow \; a_1 \: + \: M \: + \: a_2,
\end{equation}
convoluted with the probabilities to find partons $a_1, a_2$ in
the incoming protons.  The process is shown in Fig.~5, where the
probabilities are denoted by the effective parton densities $G
(x_i)$. At the parton-parton level, (\ref{eq:b7}), the
unintegrated gluon distributions in partons $a_1, a_2$ may be
calculated perturbatively in terms of the non-forward BFKL
amplitudes $A_1, A_2$.  This means that we must replace the
densities $f_g$ in (\ref{eq:a4}) by $A_i \sqrt{T_i}$, where $T_i
\equiv T (k_{it}, \mu)$ is given by (\ref{eq:a7}).  The
non-forward BFKL amplitudes $A_i$ are of the form \cite{FR}
\begin{equation}
\label{eq:a12}
 A_i \; = \; \exp (-n_i/2) \: \Phi (Y_i),
\end{equation}
where $n_i$ is the mean number of gluons emitted, with transverse
momenta in the range $(Q_t, k_{it})$ and rapidities in the
interval $\Delta \eta_i$
\begin{equation}
\label{eq:a13}
 n_i \; = \; \frac{3 \alpha_S}{\pi} \: \Delta \eta_i \: \ln \left
 ( \frac{k_{it}^2}{Q_t^2} \right ).
\end{equation}
The remaining factor $\Phi (Y_i)$ accounts for the usual BFKL
(single) logarithms.  For rapidity gaps $\Delta \eta_i < 4$ we
have
\begin{equation}
\label{eq:a14}
 Y_i \; \equiv \; \frac{3 \alpha_S}{2 \pi} \: \Delta \eta_i \;
 \lapproxeq \; 0.4,
\end{equation}
and, for $Q_t^2 \ll k_{it}^2$, it is sufficient to retain only the
$O (Y_i)$ term, which gives \cite{FR}
\begin{equation}
\label{eq:a15}
 \Phi (Y_i) \; \simeq \; 1 \: + \: Y_i \: \frac{Q_t^2}{k_{it}^2}
 \; \simeq \; 1.1 \: \pm \: 0.1.
\end{equation}
We use this value in our numerical predictions.  Note that the
BFKL amplitudes, which describe elastic parton-parton scattering,
already contain the double-logarithmic Sudakov suppression which
reflects the absence of secondary gluon emission with transverse
momenta $p_t$ up to the momentum transfer $k_{ti}$.  The remainder
of the hard suppression factor, which accounts for the absence of
emission in the interval $(k_{ti}, \mu)$, has been incorporated
separately in terms of the survival factors $T_i$.

The dominant (leading $\log$) contribution to the inclusive
amplitude of Fig.~1(b) comes from the asymmetric configuration
$Q_t^2 \ll k_{it}^2$.  Thus the transverse momenta of the two hard
active gluons can no longer be approximated by $k_{it} \approx
Q_t$, as they were in the exclusive case in (\ref{eq:a4})
\cite{KMR}.  As a consequence each hard gluon propagator, together
with its polarization factor, can no longer be written as
$Q_t/Q_t^2$, but must remain $k_{it}/k_{it}^2$.  Therefore, the
factor $1/Q_t^4$ in the amplitude in (\ref{eq:a4}) now becomes
$1/(Q_t^2 k_{1t} k_{2t})$.  Moreover, in the limit $Q_t^2 \ll
k_{it}^2$ we have $k_{it} \simeq k_{it}^\prime$.  Thus, instead of
the factor $1/b$, which came from the $t_i$ integral limited by
the proton form factor as in (\ref{eq:a4}) and (\ref{eq:a5}), we
now obtain the logarithmic $dk_{it}^2/k_{it}^2$ integrals.  Here
we have used
\begin{equation}
\label{eq:a17}
 t_i \; = \; (Q - k_i)^2 \; \simeq \; - k_{it}^2 \; \simeq \;
 -k_{it}^{\prime 2}.
\end{equation}
Thus, to compute Fig.~5 we must evaluate
\begin{equation}
\label{eq:a9}
 {\cal I} \; = \; \int \: \frac{dQ_t^2}{Q_t^2} \: \frac{dQ_t^{\prime 2}}{Q_t^{\prime
 2}} \: \frac{dk_{1t}^2}{k_{1t}^2} \: \frac{dk_{2t}^2}{k_{2t}^2}
 \: (A_1 A_2 A_1^\prime A_2^\prime) \: \sqrt{T_1 T_1^\prime T_2
 T_2^\prime}.
\end{equation}

In this way, we find the luminosity, (\ref{eq:a3}), for the
inclusive process of Fig.~1(b) and Fig.~5, in which the initial
protons dissociate and the system $M$ is produced with rapidity
gaps $\Delta \eta_1$ and $\Delta \eta_2$ on either side, is, to
leading $\log$ accuracy, expressed in the form \cite{KMRdijet}
\begin{equation}
\label{eq:a8}
 L^{\rm incl} \; = \; \int_{x_1^{\rm min}}^1 \: G (x_1)
\: \frac{dx_1}{x_1}\: \int_{x_2^{\rm min}}^1 \: G (x_2) \:
\frac{dx_2}{x_2} \: \frac{\alpha_S^4}{\pi^2} \: \left
(\frac{N_C^2}{N_C^2 - 1} \right )^2 \: {\cal I},
\end{equation}
with $N_C = 3$.  The primed quantities arise because the
luminosity is obtained by multiplying the inclusive amplitude by
its complex conjugate, as in Fig.~5.  The effective parton
densities,
\begin{equation}
\label{eq:a10}
 G (x_i) \; = \; x_i g (x_i, k_{it}^2) \: + \: \frac{16}{81} \:
 \sum_q \: x_i \left ( q (x_i, k_{it}^2) \: + \: \bar{q} (x_i,
 k_{it}^2) \right ),
\end{equation}
are integrated from
\begin{equation}
\label{eq:a11}
 x_i^{\rm min} \; = \; \frac{M}{\sqrt{s}} \: e^y \: + \:
 \frac{k_{it}}{\sqrt{s}}\: e^{(y + \Delta \eta_i)}
\end{equation}
up to 1.  These limits ensure that no recoil jets in the
dissociated states $X$ and $Y$ lie within the rapidity gap
intervals $\Delta \eta_1$ and $\Delta \eta_2$ respectively.

Due to the asymmetric configurations of the $t$-channel gluons,
$Q_t \ll k_{it}$, we have, besides $\Delta \eta_i$, a second
logarithm $\ln (k_{it}^2/Q_t^2)$ in the BFKL evolution.  Using
this double $\log$ result (that is setting $\Phi (Y_i) = 1$ in
(\ref{eq:a12})), we may simplify expression (\ref{eq:a9}) for
${\cal I}$. Then we perform the $Q_t^2$ and $Q_t^{\prime 2}$
integrations and find
\begin{equation}
\label{eq:a16}
 {\cal I} \; = \; \frac{1}{(Y_1 + Y_2)^2} \: \int \:
 \frac{dt_1}{t_1} \: \frac{dt_2}{t_2} \: \exp \left ( - \:
 \frac{3 \alpha_S}{\pi} \: \Delta \eta \: \left | \ln
 \frac{t_1}{t_2}\right | \right ) \; T \left ( \sqrt{|t_1|}, \mu
 \right ) \: T \left (\sqrt{|t_2|}, \mu \right ),
\end{equation}
where $\Delta \eta$ equals $\Delta \eta_1$ if $|t_1| > |t_2|$, but
equals $\Delta \eta_2$ when $|t_1| < |t_2|$.

The luminosity $M^2 d{\cal L}/dy dM^2$ calculated from
(\ref{eq:a8}) and (\ref{eq:a3}) for inclusive double-diffractive
production of a system of mass $M$ and rapidity $y$ is shown in
Figs.~2 and 3, by the thin continuous curves corresponding to two
choices of rapidity gaps, namely $\Delta \eta_1 = \Delta \eta_2 =
2$ and, $\Delta \eta_1 = \Delta \eta_2 = 3$ in Fig.~1(b).  Again
we use MRST99 partons \cite{MRST}, to calculate (\ref{eq:a10}).
Since we are working at the partonic level, a Monte Carlo
simulation would be required to obtain the precise experimental
prediction. Unfortunately, the lower curve for $\Delta \eta = 3$
is more relevant, as after hadronization the smaller rapidity gaps
with $\Delta \eta = 2$ may be hard to identify.

The effective luminosity, that is the number of $gg$ collisions,
is larger for inclusive than exclusive production due to the
larger available phase space, since now the transverse momentum of
the heavy system $M$ is no longer limited by the proton form
factor. Moreover partons of larger $k_{ti}$ tend to have large
momentum fractions $x_i$, see (\ref{eq:a11}).  For partons of
relatively large $x_i$ we sample mainly the smaller size component
of the proton wave function, which in turn has the smaller
absorption cross section \cite{KKMR}.  This leads to a larger
`soft' survival probability $\hat{S}^2$.  To be precise we use
model II of Ref.~\cite{KKMR} to calculate $\hat{S}^2$.  For
inclusive production, the survival factor $\hat{S}^2$ depends on
the values of $M^2$ and $\Delta \eta$, since these alter the
values of $x_i$ sampled.

\subsection{Production via Pomeron-Pomeron fusion}

The inclusive process studied in the previous subsection allowed
secondary particles in the proton fragmentation regions, see
Fig.~1(b).  An alternative possibility is to detect both
elastically scattered protons, but to permit secondaries in the
central rapidity interval $\delta \eta$ containing the heavy
system, as schematically sketched in Fig.~1(c).  In lowest order
perturbative QCD, this process is described by Fig.~6(a), which is
similar to Fig.~1(a) for exclusive production.  The only
difference is that now we have bremsstrahlung associated with the
hard subprocess $gg^{PP} \rightarrow M$.  Therefore we can compute
the cross section from the same formula (\ref{eq:a4}), except that
the Sudakov-like suppression is weaker.  Now the $1 - \Delta$
upper limit in the $z$ integration in (\ref{eq:a7}) is specified
by
\begin{equation}
\label{eq:b17}
 \Delta \; = \; \left ( \frac{k_t}{\mu + k_t} \right ) \: \cosh
 \left ( \frac{\delta \eta}{2} \right ).
\end{equation}
Clearly the luminosity is larger and increases with $\delta \eta$,
as is seen from the dashed curves in Figs.~2, 3 and 4.  We label
the curves\footnote{Actually we have included the luminosity of
the exclusive process of Fig.~1(a) in the $C$-{\it inel.} results.
Strictly speaking, therefore, a more precise notation might have
been $C$-{\it total}, to indicate that the `elastic' contribution
is included.} $C$-{\it inel.} to indicate {\it central-inelastic}
double-diffractive production of the heavy system, see Fig.~6(a).
As can be seen from the figures, we do not gain much luminosity by
allowing extra emission in the central region with $\delta \eta =
2$.  If we enlarge $\delta \eta$ up to 4 we enhance the luminosity
by an order of magnitude, but at a great price, as will now be
discussed.

\begin{figure}[!h]
\begin{center}
\epsfig{figure=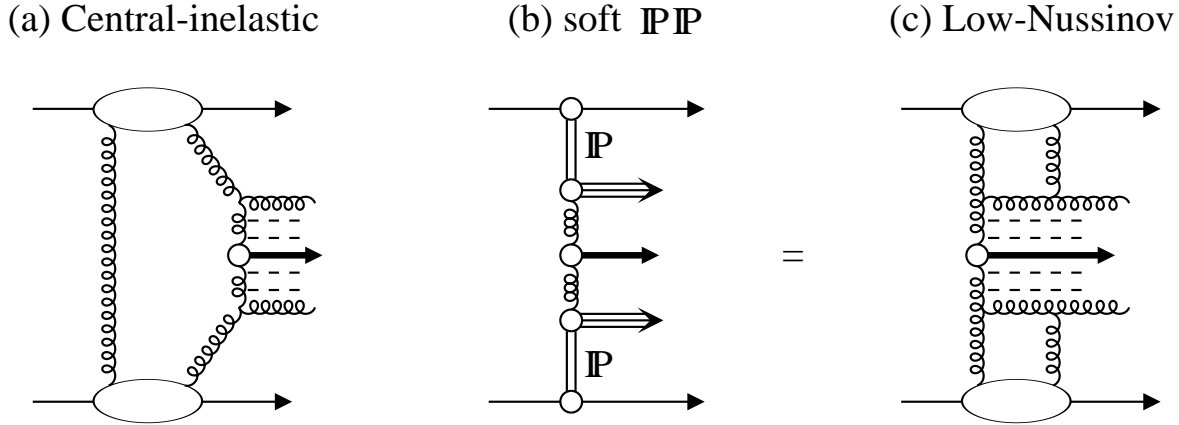,height=2.5in}
 \caption{Double-diffractive production of a heavy system, shown by
 the {\it bold} central arrow, accompanied by secondaries in the central
 region, with forward going protons.}
 \label{Fig6}
\end{center}
\end{figure}

First, it is important to note that for processes of the type of
Fig.~1(c), the mass $M$ determined by detection of the forward
protons is no longer equal to the mass of the heavy system. In
this subsection we denote the former mass by $M_{PP}$ and the
latter by $M_H$.  We see that $M_{PP} > M_H$ due to the presence
of secondaries in the central region.  This clearly applies to the
process of Fig.~6(a) which we have just discussed, but also to the
soft $\funp\funp$ fusion process of Fig.~6(b).  For these
inelastic $\funp\funp$ processes we loose all the advantages of
the exclusive process.  The clean central environment of the heavy
system is now populated by secondaries.  The mass $M_H$ can
therefore no longer be measured by the missing-mass method.
Moreover, there is no $J_z = 0$ selection rule to suppress the QCD
background.

Since the soft Pomeron-Pomeron fusion process of Fig.~6(b) is at
present frequently discussed \cite{CF,PR}, we study it in more
detail to confirm that it is indeed not a viable search mechanism.
In terms of Feynman graphs, Fig.~6(b) corresponds to Fig.~6(c) in
which the soft Pomerons have been replaced by two-gluon exchange,
as originally considered by Low \cite{L} and Nussinov \cite{N}.
Interestingly, we see that the cross section for the process in
Fig.~6(c) is suppressed by a factor $\alpha_S^2$ in comparison to
that of Fig.~6(a).  In the cross section for Fig.~6(b) where,
following Ref.~\cite{IS}, the Pomerons are treated as real
particles with their own parton distributions, the factor is
reflected in the small values of the effective Pomeron structure
functions which are determined phenomenologically from diffractive
data.  It therefore may not be quite so surprising when we show
below that the soft Pomeron-Pomeron fusion process of Fig.~6(b)
has a {\it smaller} cross section than exclusive
double-diffractive production of Fig.~1(a). The cross
section\footnote{The cross section was also calculated in
\cite{PR}, where a larger value was found, which we cannot
reproduce.} for soft Pomeron-Pomeron fusion was also recently
discussed in Ref.~\cite{CF}.

The luminosity factor for the soft Pomeron-Pomeron fusion process,
shown in Fig.~6(b), is
\begin{equation}
\label{eq:a18}
 L^{\rm soft} \; = \; \left ( \frac{\sigma_0}{16 \pi^2} \right )^2
 \: \int \: dt_1 dt_2 \: F_N^2 (t_1) \: F_N^2 (t_2) \: \left (
 \frac{1}{x_1} \right )^{2\left(\alpha (t_1) - 1\right)} \: \left (
 \frac{1}{x_2} \right )^{2\left(\alpha (t_2) - 1\right)},
\end{equation}
where we assume the Donnachie-Landshoff parametrization of the
elastic $pp$ amplitude \cite{DL}
\begin{eqnarray}
\label{eq:a19}
 A_{pp} (s, t) & = & i \sigma_0 \: F_N^2 (t) \: s^{\alpha (t)} \nonumber \\
 & & \nonumber \\
 {\rm Im}~A (s, t = 0) & = & s \: \sigma_{\rm tot} (s) \\
 & & \nonumber \\
 \alpha (t) & = & 1.08 \: + \: 0.25~t \nonumber
\end{eqnarray}
with $s$ and $t$ in units of GeV$^2$.  $F_N$ is the proton form
factor and $\sigma_0 = 21.7$~mb.  The mass and rapidity of the
produced system are given by
\begin{equation}
\label{eq:b19}
 M_{PP}^2 \; = \; x_1 \: x_2 \: s, \quad\quad\quad y \; = \;
 \frac{1}{2} \: \ln \left ( \frac{x_1}{x_2} \right ).
\end{equation}

For this production mechanism the cross section $\hat{\sigma}$ in
(\ref{eq:a2}), which multiplies the luminosity, is the standard
inelastic cross section given by the convolution of the parton
distributions (of the Pomerons) $a_i^\funp$ with the cross section
for the $a_1 a_2 \rightarrow M$ hard subprocess
\begin{equation}
\label{eq:c19}
 \hat{\sigma} \; = \; \sum_{a_1, a_2} \: \int \: \frac{dz_1}{z_1}
 \: \int \frac{dz_2}{z_2} \: z_1 a_1^\funp (z_1) \: z_2
 a_2^\funp (z_2) \: \sigma (a_1 a_2 \rightarrow M),
\end{equation}
with $a = g, q$.  To estimate $\hat{\sigma}$ we assume gluon
dominance and that $zg^\funp (z) \simeq 0.7$ for our kinematic
domain, which is consistent with the distribution obtained by the
H1 collaboration from the analysis of their diffractive data
\cite{H1}.  Then (\ref{eq:c19}) gives
\begin{eqnarray}
\label{eq:d19}
 \hat{\sigma} & \simeq & (0.7)^2 \: \sigma (gg \rightarrow M) \:
 \int \: \frac{dz_1}{z_1} \: \int \: \frac{dz_2}{z_2} \nonumber \\
 & & \\
 & \simeq & (0.7)^2 \: \sigma (gg \rightarrow M) \: (\delta
 y)^2/2, \nonumber
\end{eqnarray}
where
\begin{equation}
\label{eq:e19}
 \delta y \; = \; \ln (M_{PP}^2/M_H^2).
\end{equation}
Here, to be definite, we have assumed that the heavy system $M$ is
a Higgs boson of mass $M_H$.  Next, it is important to note that
$\sigma (gg \rightarrow H)$ for the inelastic subprocess within
the soft $\funp\funp$ fusion mechanism of Fig.~6(b) is a factor
$1/2 (N_C^2-1)$ {\it smaller} than the corresponding $\sigma (gg
\rightarrow H)$ for the exclusive process of Fig.~1(a).  Recall
that this factor arises because the exclusive process proceeds via
the {\it coherent} fusion of gluons of different spin and colour,
which leads to an enhancement of $2 (N_C^2 - 1)$ in comparison
with the inclusive process of Fig.~6(b), see (\ref{eq:b2}) and
(\ref{eq:c2}).

In addition, at Tevatron energies, after observing the leading
outgoing proton and antiproton, the available phase space $\delta
y$ becomes rather small, $\delta y \sim 1$.  In summary, the cross
section $\hat{\sigma}$ multiplying the luminosity in (\ref{eq:a2})
is a factor of about
\begin{equation}
\label{eq:f19}
 (0.7)^2 \: \frac{1}{2 (N_C^2 - 1)} \: \frac{(\delta y)^2}{2} \;
 \simeq \; \frac{1}{64}
\end{equation}
smaller for production via soft $\funp\funp$ fusion (Fig.~6(b))
than via the `clean' exclusive process of Fig.~1(a).  Now from
Fig.~3, we see that, for the production of a Higgs boson of mass
$M_H = 120$~GeV, the soft $\funp\funp$ luminosity calculated from
(\ref{eq:a18}) is only a factor of about 10 larger than the
luminosity for exclusive production.  In both processes we use the
same survival factor\footnote{Another way to estimate $\hat{S}^2$
is to compare the data with the theoretical prediction which does
not account for the soft rescattering effects.  Special care
should be taken if this method is used.  If the `bare' theoretical
cross section is calculated using, for example, only the mechanism
of Fig.~6(b) and does not include the more important central
inelastic production mechanism of Fig.~6(a), then $\hat{S}^2 =$
data/theory will be overestimated.} $\hat{S}^2$, which was given
at the end of subsection 2.1.  Thus we obtain our advertised
result:  the expected rate of double-diffractive Higgs production
at the Tevatron in the `dirty' environment of inelastic
$\funp\funp$ fusion (Fig.~6(b)) is in fact {\it
smaller}\footnote{This estimate of the soft Pomeron-Pomeron
contribution is in reasonable agreement with that of
Ref.~\cite{CF}.} than the rate for production by the `clean'
exclusive process of Fig.~1(a).

Of course, at LHC energies the available phase space $\delta y$ is
larger.  Recall, from (\ref{eq:d19}), $\hat{\sigma}$ grows as
$(\delta y)^2$.  On the other hand, in going from Tevatron to LHC
energies, the ratio of the exclusive to soft $\funp\funp$
luminosities increases by more than an order of magnitude (see
Fig.~2), due to the growth of gluons at small $x$.  We conclude
that there is nothing to be gained by studying the soft
$\funp\funp$ fusion mechanism, at least from the point of view of
Higgs searches.

\subsection{Production via $\gamma\gamma$ and $WW$ fusion}

For completeness we give the luminosity factors assuming that the
rapidity gaps are due to the production of the $(M,y)$ system via
$\gamma\gamma$ and $WW$ fusion.  We have
\begin{equation}
\label{eq:a20}
 L^{\gamma\gamma} \; = \; \left ( \frac{\alpha}{\pi} \right )^2 \:
 \int_{t_{1,\:{\rm min}}} \: dt_1 \frac{(t_1- t_{1,\:{\rm min}})}{t_1^2} \:
 \left (F^{\rm em} (t_1) \right )^2 \: \int_{t_{2,\:{\rm
 min}}} \: dt_2 \frac{(t_2 - t_{2,\:{\rm min}})}{t_2^2} \: \left (F^{\rm em} (t_2)
 \right )^2,
\end{equation}
where $|t_{i, {\rm min}}| = x_i^2 m_N^2$, $m_N$ is the proton mass
and $\alpha = 1/137$. As the dominant (logarithmically enhanced)
contribution comes from very small $|t_i|$, that is large impact
parameters, here we would expect the survival factor $\hat{S}^2$
to be close to 1 \cite{KMR}.  Indeed, we find this to be the case,
as illustrated by the values of $\hat{S}^2$ that are given in
Section~3.1.3 for Higgs production via $\gamma\gamma$ fusion.  The
$\gamma\gamma$ luminosity, (\ref{eq:a3}), shown in Fig.~2 is
calculated\footnote{Besides the proton form factor $F_1^{\rm em}$,
we also include the small contribution from $F_2^{\rm em}$, when
we evaluate $F^{\rm em} (t)$ in (\ref{eq:a20}).} from
$L^{\gamma\gamma}$, and includes the gap survival factor
$\hat{S}^2$.

The effective luminosity for producing the $(M,y)$ system by $WW$
fusion is given by
\begin{eqnarray}
\label{eq:a21}
 L^{WW} & = & \left [ \int_{x_1^{\rm min}}^1 \: U (x_1) \:
 \frac{dx_1}{x_1} \: \int_{x_2^{\rm min}}^1 \: D (x_2) \:
 \frac{dx_2}{x_2} \: + \: \int_{x_1^{\rm min}}^1 \: D (x_1) \:
 \frac{dx_1}{x_1} \: \int_{x_2^{\rm min}}^1 \: U (x_2) \:
 \frac{dx_2}{x_2} \right ] \nonumber \\
 & & \\
 & & \times \; \left ( \frac{g^2}{16 \pi^2} \right )^2 \: \int \:
 \frac{dt_1 M_W^2}{(M_W^2 - t_1)^2} \: \frac{dt_2 M_W^2}{(M_W^2 -
 t_2)^2}, \nonumber
\end{eqnarray}
where $g^2 = 8 M_W^2 G_F/\sqrt{2}$ and $M_W$ is the mass of the
$W$ boson.  The lower limits of the integrations are given by
(\ref{eq:a11}).  The effective parton densities, $U (x_i,
k_{it}^2)$ and $D (x_i, k_{it}^2)$ are
\begin{eqnarray}
\label{eq:a22}
 U (x) & = & xu \: + \: x\bar{d} \: + \:
 x\bar{s} \: + \: xc \nonumber \\
 & & \\
 D (x) & = & x\bar{u} \: + \: xd \: + \: xs \:
 + \: x\bar{c}. \nonumber
\end{eqnarray}
The survival factor $\hat{S}^2$ is calculated using the
two-channel eikonal approach described in Ref.~\cite{KKMR}. We use
the more realistic model II \cite{KKMR}, as in Section~2.2.

\begin{figure}[!h]
\begin{center}
\epsfig{figure=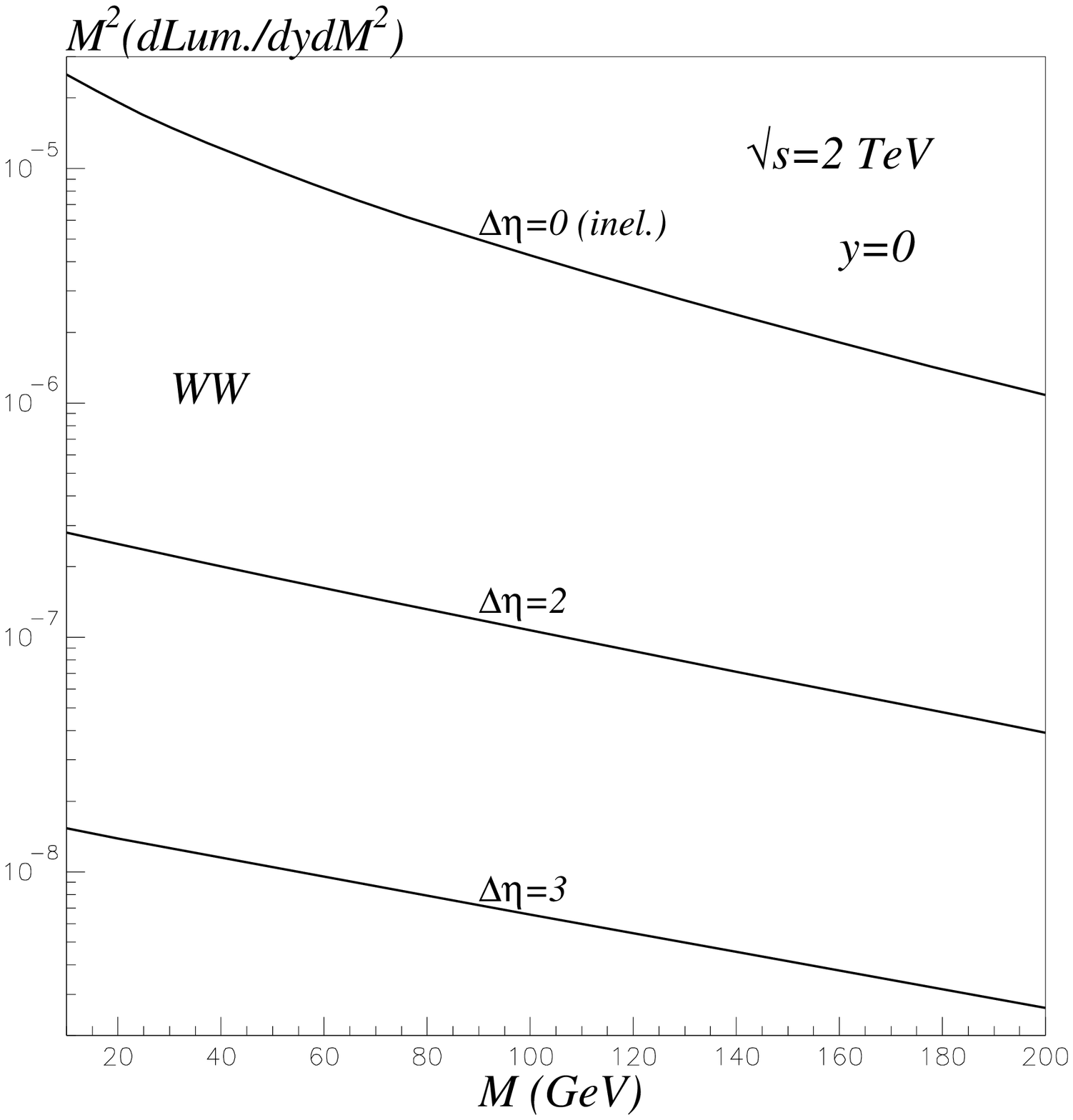,height=3.3in}
\epsfig{figure=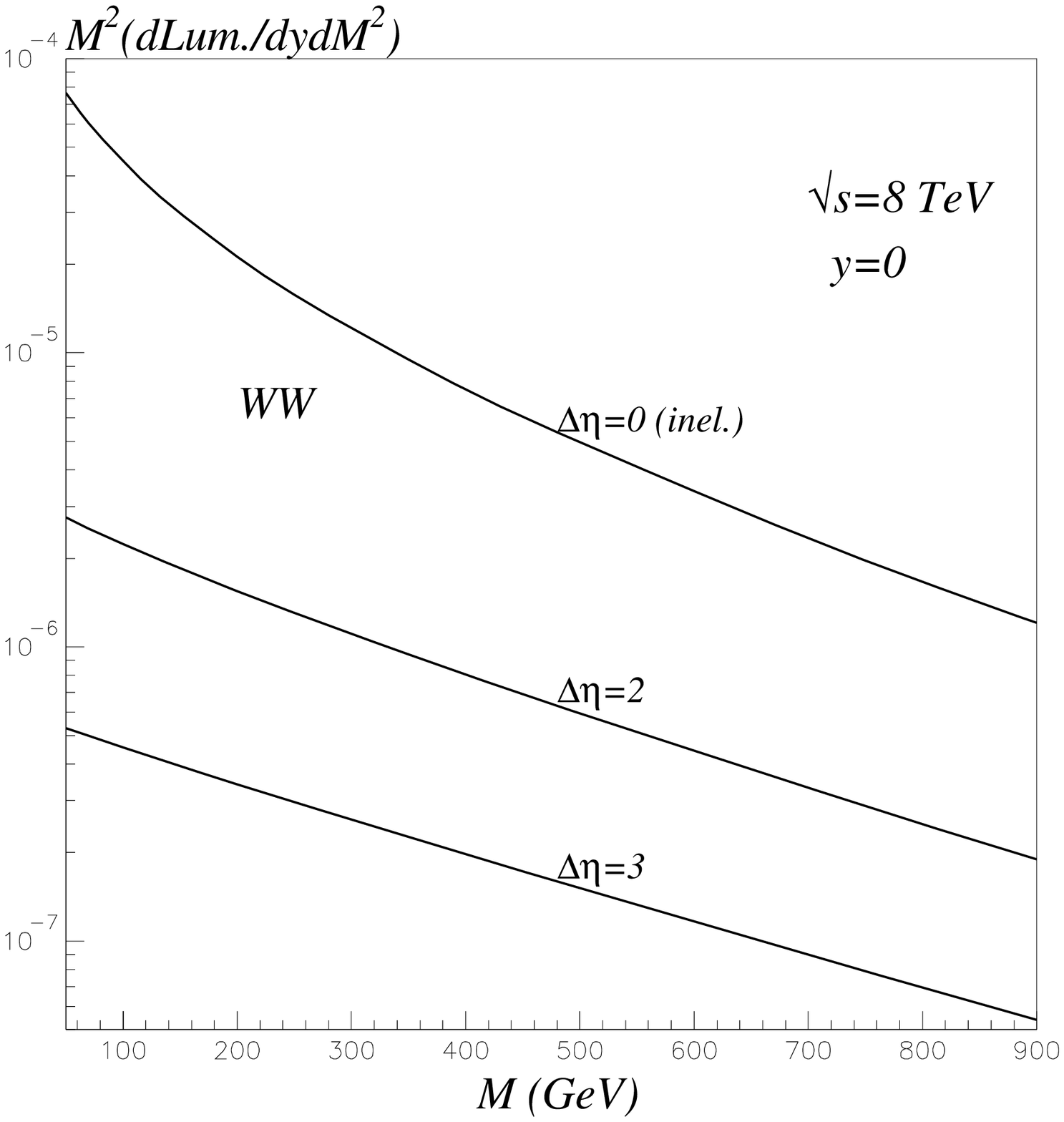,height=3.3in}
\epsfig{figure=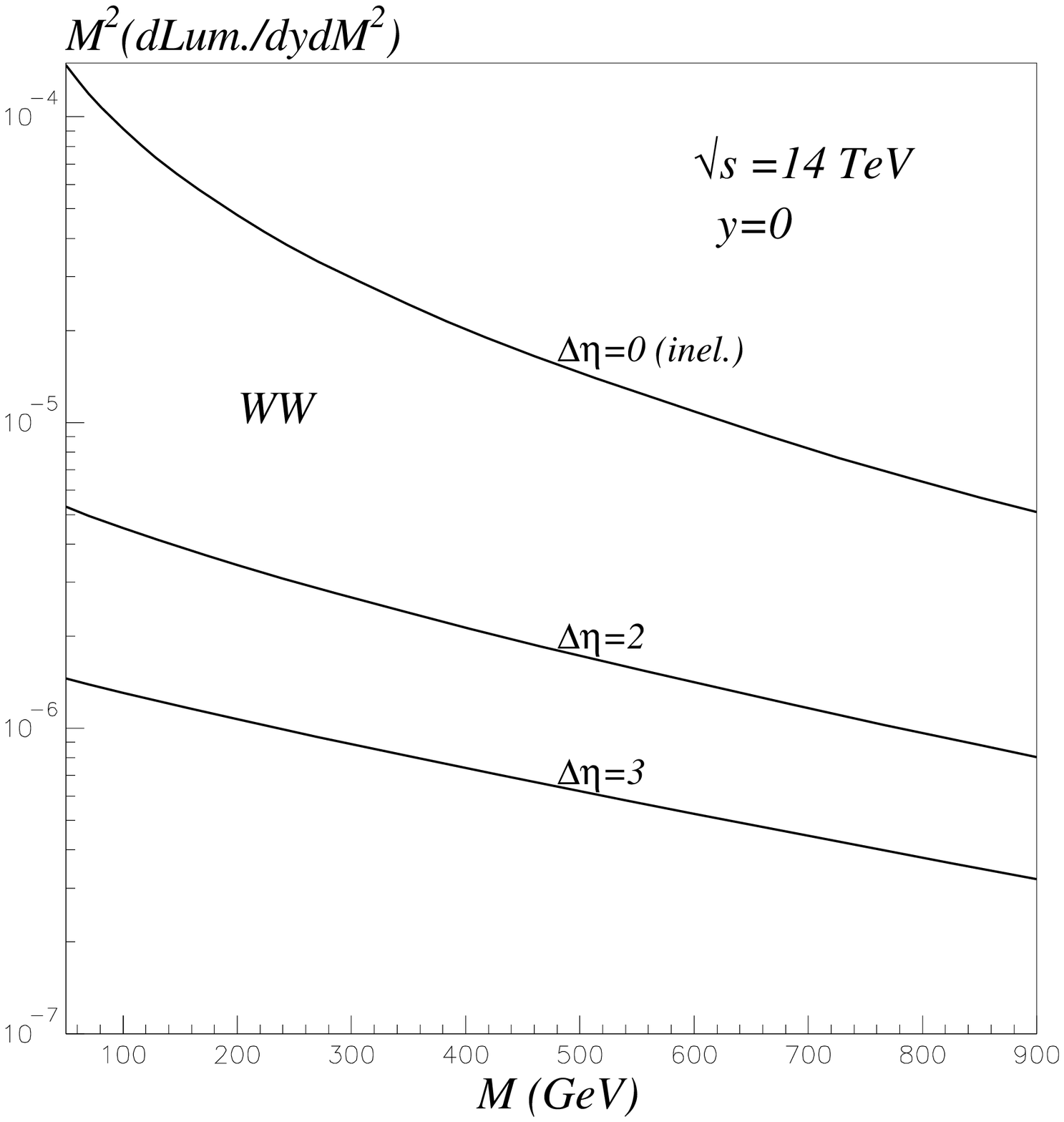,height=3.3in}
 \caption{The luminosity $M^2 \partial {\cal L}/\partial y \partial M^2$
 versus $M$ for the production of a heavy system at $y = 0$ by $WW$ fusion,
 with a rapidity gap of size $\Delta \eta$ on either side, at $pp$ or
 $p\bar{p}$ collider energies of $\sqrt{s} = 2, 8$ and 14~TeV.}
 \label{Fig7}
\end{center}
\end{figure}

Predictions for the $WW$ luminosity are shown in Figs.~3, 4 and 7.
Since $WW$ fusion can only mediate inclusive production, the
missing mass technique cannot be used.  However, here, the
enlargement of the rapidity gaps does not reduce the luminosity as
much as it did for $gg$ fusion.  Fig.~7 shows the $WW$ luminosity
for $\Delta \eta = 2$ and $\Delta \eta = 3$, in comparison with
that for conventional inelastic production via $WW$ fusion for
which $\Delta \eta = 0$.  These results should be compared with
the results for $gg$ fusion in Figs.~3 and 4.  The upper (dotted)
curve shows the inelastic luminosity $(\Delta \eta = 0)$ assuming
that the $gg$ pair are in a colour singlet state.  If this
luminosity is compared with the inclusive luminosity for the
production of a $M = 120$~GeV system with rapidity gaps of $\Delta
\eta = 3$ on either side, then we loose a factor of $10^{-3}$ at
LHC energies, while the corresponding ratio for production via
$WW$ fusion is only 0.02.  Production by $WW$ fusion is less
suppressed with increasing $\Delta \eta$ for two reasons.  First,
since the $W$ bosons originate mainly from valence quarks, that is
the component of the proton wave function with the smaller
absorption cross section, the soft survival probability of the
rapidity gaps, $\hat{S}^2$, is larger.  Secondly, as the $W$ boson
is a colourless object, there is no Sudakov suppression due to QCD
bremsstrahlung.

\section{Cross sections for the hard subprocesses and for double-diffractive production}

Selecting double-diffractive events means essentially that we are
working with a gluon-gluon collider\footnote{For completeness, we
also consider production mechanisms where the rapidity gaps are
associated with $\gamma\gamma$ and $WW$ fusion processes.}.
Indeed, the luminosities presented in Section~2 were given in
terms of the numbers of $gg$ collisions per proton-proton
interaction.  As emphasized before, there is an important
distinction between exclusive and inclusive double-diffractive
production of a heavy system $M$.  In {\it exclusive} events the
forward protons select a colour singlet, $J_z = 0$ incoming $gg$
state.  In {\it inclusive} events the $J_z = 0$ selection rule is
absent. Therefore, to predict the cross section (or event rate at
the collider) we have to be careful to ensure that we multiply the
relevant luminosity by the appropriate $gg^{PP} \rightarrow M$
subprocess cross section.  Recall that the $PP$ superscript was to
indicate that the gluons are mediating a double-diffractive
process (loosely called Pomeron-Pomeron production).

In this section we give the subprocess cross sections
$d\hat{\sigma}/d\Omega$ which are relevant to the
double-diffractive production of various heavy systems $M$.  The
differential form is symbolic.  For example, for the production of
a dijet system, with a rapidity gap on either side, we study
$d\hat{\sigma}/dE_T^2$, where $E_T$ is the transverse energy of
the jets.  In this way we can readily predict the cross sections
for double-diffractive production at various collider energies,
using
\begin{equation}
\label{eq:a23}
 M^2 \: \frac{d \sigma^{(i)}}{dy dM^2 d\Omega} \; = \; \left (M^2
 \: \frac{d {\cal L}^{(i)}}{dy dM^2} \right ) \: \frac{d\hat{\sigma}^{(i)}
 (M^2)}{d\Omega},
\end{equation}
see (\ref{eq:a2}).  The expression in brackets is the luminosity
calculated in Section~2, which is to be multiplied by the
appropriate subprocess cross section, which we enumerate below. We
study in turn, resonance production (e.g.\ Higgs, $\chi_b$), dijet
production, $\gamma\gamma$ production, $t\bar{t}$ production, the
production of SUSY particles (e.g.\ $\tilde{g}\tilde{g},
\tilde{q}\tilde{\bar q}$), and various soft phenomena.

\subsection{Resonance production}

The subprocess cross section for the double-diffractive production
of a $0^{++}$ resonance $R$ may be written in terms of its
two-gluon partial width
\begin{eqnarray}
\label{eq:a24}
 \hat{\sigma}^{\rm excl} (gg^{PP} \rightarrow R) & = & \frac{2 \pi^2 \tilde{\Gamma} (R \rightarrow
 gg)}{M_R^3} \: \delta \left ( 1 \: - \: \frac{M^2}{M_R^2} \right
 ), \\
 & & \nonumber \\
 \label{eq:a25}
 \hat{\sigma}^{\rm incl} & = & \frac{1}{2} \: \hat{\sigma}^{\rm
 excl}.
\end{eqnarray}
The extra factor of $\frac{1}{2}$ arises in inclusive
double-diffractive production due to the absence of the $J_z = 0$
selection rule in this case.  It reflects the spin coherence of
the {\it exclusive} process where the incoming gluon polarisations
are correlated, whereas for {\it inclusive} production, the cross
section is averaged over the incoming spins in the usual way, that
is
\begin{equation}
\label{eq:a26}
 \hat{\sigma}^{\rm excl} \; \sim \; |\overline{\cal M}|^2,
 \quad\quad\quad \hat{\sigma}^{\rm incl} \; \sim \; \overline{| {\cal
 M}|^2}.
\end{equation}

\subsubsection{Higgs production}

In the Born approximation, the width $\tilde{\Gamma}$ in
(\ref{eq:a24}) is just the lowest order $R \rightarrow gg$ decay
width $\Gamma_0$.  For Higgs production we include the NLO
correction to $\hat{\sigma} (gg \rightarrow H)$, using
\begin{eqnarray}
\label{eq:a27}
 \tilde{\Gamma} (H \rightarrow gg) & = & \Gamma_0 (H \rightarrow
 gg) \: \left ( 1 \: + \: \frac{\alpha_S (M_H)}{\pi} \: \left (
 \pi^2 \: + \: \frac{11}{2} \right ) \right ) \nonumber \\
 & & \\
 & \simeq & 1.5 \: \Gamma_0 (H \rightarrow gg). \nonumber
\end{eqnarray}
Note that (\ref{eq:a27}) is not the complete NLO correction to the
$H \rightarrow gg$ partial width, since it does not include the
contribution from the emission of QCD radiation, see, for example,
\cite{MS,ZK}. For a Higgs boson of mass $M_H = 120$~GeV, we obtain
from (\ref{eq:a24}) and (\ref{eq:a27})
\begin{equation}
\label{eq:a28}
 \hat{\sigma}^{\rm excl} \; \simeq \; \delta \left ( 1 \: - \:
 \frac{M^2}{M_H^2} \right ) \: 1.1~{\rm pb}.
\end{equation}
In the intermediate mass range of the Higgs, $\hat{\sigma}$
depends weakly on $M_H$, as $\Gamma_0 \sim M_H^3 \alpha_S^2 (M_H)$
largely cancels the $M_H^{-3}$ factor in (\ref{eq:a24}).

Given the subprocess cross section (\ref{eq:a28}) and the
luminosities of Figs.~2, 3 and 4, we can readily predict the cross
section for double-diffractive Higgs production at the LHC and the
Tevatron.  To be specific let us take $M_H = 120$~GeV.  Then from
Fig.~3 we have
\begin{eqnarray}
\label{eq:a29}
 \left . \frac{d {\cal L}}{dy} \right |_{y = 0} & \simeq & 0.6 \:
 \times \: 10^{-3} \; ({\rm LHC}) \nonumber \\
 & & \\
 & \simeq & 0.9 \: \times \: 10^{-4} \; ({\rm Tevatron}). \nonumber
\end{eqnarray}
Also we see that for Higgs production $\Delta y \simeq 5$ (LHC) or
2 (Tevatron) will allow a quick estimate of the $y$ integration.
Finally multiplying by $\hat{\sigma}$ of (\ref{eq:a28}) (which
takes care of the $M^2$ integration), we obtain
\begin{eqnarray}
\label{eq:a30}
 \sigma (pp \rightarrow p + H + p) & \simeq & 3~{\rm fb} \; ({\rm LHC})
 \nonumber \\
 & & \\
 \sigma (p\bar{p} \rightarrow p + H + \bar{p}) & \simeq & 0.2~{\rm fb} \;
 ({\rm Tevatron}). \nonumber
\end{eqnarray}
These predictions are consistent with our previous estimates
\cite{KMR}.  The background to this signal for the Higgs boson is
discussed in Section~3.2.3.

\subsubsection{$\chi_b$ production}

Another interesting example is $\chi_b (0^{++})$ resonance
production, see, for example, \cite{JPU,KMRmm,FENG}.  Again
$\hat{\sigma}$ is given by (\ref{eq:a24}), with \cite{CHIC}
\begin{equation}
\label{eq:a31}
 \tilde{\Gamma} (\chi_b \rightarrow gg) \; = \; \Gamma_0 (\chi_b
 \rightarrow gg) \: \left (1 + 9.8 \: \frac{\alpha_S}{\pi} \right
 ) \; = \; 550~{\rm keV},
\end{equation}
where we have used the lattice result $\Gamma_0 = 354$~keV
\cite{CHIW}. This gives
\begin{equation}
\label{eq:a32}
 \hat{\sigma}^{\rm excl} \; \simeq \; \delta \left ( 1 \: - \:
 \frac{M^2}{M_\chi^2} \right ) \: 3.8~{\rm nb}.
\end{equation}
As an example, consider $\chi_b$ production at the Tevatron.  From
Fig.~2 we have
\begin{equation}
\label{eq:a33}
 \left . \frac{d {\cal L}}{dy} \right |_{y = 0} \; \simeq \; 1.6
 \: \times \: 10^{-2}, \quad\quad\quad \Delta y \; \simeq \; 2,
\end{equation}
which combined with (\ref{eq:a32}) gives
\begin{equation}
\label{eq:a34}
 \sigma (p\bar{p} \rightarrow p + \chi_b + \bar{p}) \; \simeq \;
 120~{\rm pb},
\end{equation}
consistent with the predictions of Ref.~\cite{KMRmm}.

\subsubsection{Higgs production via $\gamma\gamma$ and $WW$ fusion}

For completeness we estimate Higgs production by $\gamma\gamma$
and $WW$ fusion.  For $\gamma\gamma$ fusion, (\ref{eq:a24}) is
replaced by
\begin{equation}
\label{eq:a35}
 \hat{\sigma} (\gamma\gamma \rightarrow H) \; = \;
 \frac{8 \pi^2 \tilde{\Gamma} (H \rightarrow
 \gamma\gamma)}{M_H^3} \: \delta \left ( 1 \: - \:
 \frac{M^2}{M_H^2}\right ).
\end{equation}
The factor of 8, in place of 2, arises because for coloured gluons
there was an extra factor of $(N_C^2 - 1)^{-1}$ due to colour
averaging and the absence of the $J_z = 0$ selection rule.  For
$M_H = 120$~GeV we have $\tilde{\Gamma} (H \rightarrow
\gamma\gamma) \simeq 7.9$~keV, and so
\begin{equation}
\label{eq:a36}
 \hat{\sigma} (\gamma\gamma \rightarrow H) \; \simeq \;
 \delta \left (1 \: - \: \frac{M^2}{M_H^2} \right ) \: 0.1~{\rm pb}.
\end{equation}
The $\gamma\gamma$ luminosity, integrated over $y$, is $1.0 \times
10^{-4}$ and $1.1 \times 10^{-3}$ at the Tevatron and LHC
respectively, which leads to the cross sections
\begin{eqnarray}
\label{eq:a37}
 \sigma (p\bar{p} \rightarrow p + H +
 \bar{p})_{\gamma\gamma} \; \simeq \; 0.01~{\rm fb} & & ({\rm
 Tevatron}) \nonumber \\
 & & \\
 \sigma (pp \rightarrow p + H + p)_{\gamma\gamma} \; \simeq \;
 0.1~{\rm fb} & & ({\rm LHC}). \nonumber
\end{eqnarray}
Recall that the $\gamma\gamma$ luminosity includes the gap
survival factor $\hat{S}^2$, which we calculate to be 0.75 and 0.9
at Tevatron and LHC energies respectively for the production of a
Higgs of mass $M_H = 120$~GeV.  The cross sections (\ref{eq:a37})
of the present, more detailed, calculation are lower than previous
results which were derived in the leading $\log$ approximation
with $\hat{S}^2 = 1$ \cite{EP,KMR,KMRhels}.

Note that the strong and electromagnetic contributions to
exclusive Higgs production (with rapidity gaps) have negligible
interference, because they occur at quite different values of the
impact parameter. Also, it is worth mentioning that the
$\gamma\gamma \rightarrow H$ fusion mechanism provides a natural
lower limit for the Higgs exclusive production rate.

For Higgs production via $WW$ fusion
\begin{eqnarray}
\label{eq:a38}
 \hat{\sigma}^{\rm incl} (WW \rightarrow H) & = & \frac{\pi g^2
 M_W^2}{M_H^4} \: \delta \left ( 1 \: - \: \frac{M^2}{M_H^2}
 \right ) \nonumber \\
 & & \\
 & \simeq & \delta \left ( 1 \: - \: \frac{M^2}{M_H^2} \right ) \:
 16~{\rm nb} \nonumber
\end{eqnarray}
for $M_H = 120$~GeV.  The coupling $g^2 = 8M_W^2 G_F/\sqrt{2}$.

Of course, $WW$ fusion only mediates inclusive Higgs production.
Note that there is almost no interference between the $WW$ and
$gg^{PP}$ fusion mechanisms.  Recall that the former process
produces Higgs with high transverse momentum, and as such may give
a viable signal \cite{RNC,ZEP,KMRhpt}.  Moreover, if we select
events with rapidity gaps we suppress the QCD $b\bar{b}$
background, while paying less price for the gaps in $WW$ fusion as
compared to those in $gg$ fusion (as discussed at the end of
Section 2.4).

\subsection{Dijet production}

The Born cross sections, for {\it colour-singlet} production of a
dijet system of mass $M$, are (see, for example,
\cite{BC,FKM,BL,KMRdijet})
\begin{eqnarray}
\label{eq:a39}
 \frac{d\hat{\sigma}^{\rm excl}}{dt} (gg^{PP} \rightarrow gg) & =
 & \frac{9}{4} \: \frac{\pi \alpha_S^2}{E_T^4}, \\
 & & \nonumber \\
 \label{eq:a40}
 \frac{d\hat{\sigma}^{\rm excl}}{dt} (gg^{PP} \rightarrow
 q\bar{q}) & = & \frac{\pi \alpha_S^2}{6 E_T^4} \:
 \frac{m_q^2}{M^2} \: \beta^2, \\
 & & \nonumber \\
 \label{eq:a41}
 \frac{d\hat{\sigma}^{\rm incl}}{dt} (gg^{PP} \rightarrow gg) & =
 & \frac{9}{2} \: \frac{\pi \alpha_S^2}{E_T^4} \: \left (1 \: - \:
 \frac{E_T^2}{M^2} \right )^2, \\
 & & \nonumber \\
 \label{eq:a42}
 \frac{d\hat{\sigma}^{\rm incl}}{dt} (gg^{PP} \rightarrow
 q\bar{q}) & = & \frac{\pi \alpha_S^2}{E_T^2 M^2} \: \frac{1}{6}
 \left [\left ( 1 \: - \: \frac{2 E_T^2}{M^2} \right ) \left (1 \: - \:
 \frac{2m_q^2}{E_T^2} \right ) \: + \: \frac{m_q^2}{E_T^2} \: (1 + \beta^2) \right ],
\end{eqnarray}
where $E_T$ is the transverse energy of the jets and $\beta =
\sqrt{1 - 4m_q^2/M^2}$. We choose the scale of $\alpha_S$ to be
$M/2$ to be consistent with the convention we have used for the
luminosities in Section~2. For $M^2 \gg E_T^2$ we see, that
contrary to $0^{++}$ resonance production (\ref{eq:a25}), $gg$
production is a factor 2 larger in the inclusive as compared to
the exclusive process.  This is the result of averaging over the
polarisations of the incoming gluons, (\ref{eq:c2}).  Moreover we
see that the Born cross section (\ref{eq:a40}) for exclusive
$gg^{PP} \rightarrow q\bar{q}$ production vanishes as $m_q
\rightarrow 0$ \cite{BL,PUMP}, which is a consequence of the $J_z
= 0$ selection rule for the gluon polarizations
\cite{KMRmm,KMRdijet}.

Rather than working in terms of the subprocess variables $M^2$ and
$t$, it is often convenient to use $E_T^2$ and $\delta \eta_j$,
where $\delta \eta_j = \eta_1 - \eta_2$ is the difference in
rapidities of the two outgoing jets.  It turns out that factors in
the Jacobian cancel, so that we obtain the relatively simple
relation \cite{KMRdijet}
\begin{equation}
\label{eq:a43}
 \frac{d\hat{\sigma}}{dt} \biggr / \frac{dM^2}{M^2} \; = \;
 \frac{d \hat{\sigma}}{dE_T^2} \biggr / d (\delta \eta_j).
\end{equation}
Thus the product of the luminosity (shown in Figs.~2, 3 and 4) and
$d\hat{\sigma}/dt$, will give an evaluation of the differential
cross section $d \sigma/dy d(\delta \eta_j) dE_T^2$.

The large cross section for exclusive dijet production provides
the first opportunity of experiment to check the
double-diffractive predictions, already at the Tevatron.  For
example, let us calculate the event rate for jets in the domain
defined by the intervals $25 < E_T < 35$~GeV and $\delta \eta_j <
2$. If we integrate (\ref{eq:a39}) over these intervals, and
multiply by the luminosity $\partial {\cal L}/\partial y \simeq 3
\times 10^{-4}$ appropriate to $M = 2 E_T \cosh (\delta \eta_j/2)
\simeq 70$~GeV, then we obtain
\begin{equation}
\label{eq:a44}
 \Delta \sigma (p\bar{p} \rightarrow p + jj + \bar{p}) \; \simeq
 \; 60~{\rm pb}
\end{equation}
corresponding to jets with $25 < E_T < 35$~GeV and $|\eta_1 -
\eta_2| < 2$, which is in qualitative agreement with previous
results \cite{KMRdijet,KMR}. Thus even for a Tevatron luminosity
of $1~{\rm fb}^{-1}$, we estimate there should be 60,000 exclusive
events in this particular domain.

\subsubsection{Dijets as a luminosity monitor}

There are, of course, uncertainties in our estimates of the cross
sections for double-diffractive processes.  First, the calculation
of the soft survival factor $\hat{S}^2$ is based on a simplified
two-channel eikonal model \cite{KMRsoft,KKMR}.  With the presently
available soft diffractive data, these estimates of $\hat{S}^2$
are as good as possible, but it is hard to guarantee the precision
of predictions which rely on soft physics.  Secondly recall, from
Section~2, that the luminosity was calculated in terms of the
unintegrated, skewed gluon distribution $f_g$, and that ${\cal L}
\propto f_g^4$.  Thus a 20\% uncertainty in $f_g$ results in a
factor of 2 uncertainty in the luminosity.  We calculated $f_g$ in
terms of the conventional gluon distribution obtained from the
results of a global parton analysis of deep inelastic and related
hard scattering data \cite{MRST}.  However the prescription is
only justified at leading order and for small $x$ \cite{SGMR,MR}.
In the present paper we use an improved formula for $f_g$
\cite{MR}, and as a result the double-diffractive luminosity at
Tevatron energies become a factor of two larger than in
Ref.~\cite{KMR}\footnote{The unintegrated skewed gluon
distribution $f_g (x, x^\prime, Q_t^2, \mu^2)$ has been tested,
both by providing a successful description of the data on
diffractive vector meson production at HERA \cite{MRT} and by
agreement with the upper bound on diffractive dijet production at
the Tevatron \cite{TER,BBB}.  However in both cases a much lower
scale is probed and we did not need precision better than
20--30\%.}. However the same luminosity determines the cross
sections for different double-diffractive processes. Thus dijet
production, where the cross section is largest, offers an
excellent way to monitor the $gg^{PP}$ luminosity \cite{KMR}.

\subsubsection{Dijets as a gluon factory}

Another application of the large rate of exclusive dijet
production is as a gluon factory.  The remarkable purity of the
colour-singlet, $J_z = 0$, di-gluon system provides a unique
environment to study high energy gluon jets \cite{KMRmm}.

\subsubsection{Dijets as a background to the Higgs signal}

The large dijet production rate may produce a huge background to
the Higgs $H \rightarrow b\bar{b}$ signal.  Recall that the cross
section for exclusive double-diffractive Higgs production was
calculated in Section~3.1.1.

In fact, for any $0^{++}$ resonance the ratio of the exclusive $R
\rightarrow gg$ signal to the $gg$ background is
\begin{equation}
\label{eq:a45}
 \frac{S_{gg}}{B_{gg}} \; = \; \frac{4 \pi}{9 (9.7) \: \alpha_S^2 \: \Delta
 M} \: \Gamma \left (R \rightarrow gg^{J_z = 0} \right ),
\end{equation}
where $\Delta M$ is the experimental missing-mass resolution, the
scale of $\alpha_S$ is $M/2$ and the factor 9.7 comes from
imposing a cut $60^\circ < \theta^* < 120^\circ$ in order to
improve the signal-to-background ratio.  $\theta^*$ is the decay
angle in the dijet rest frame.  Note that the major part of NLO
effects and uncertainties in the luminosity cancel in ratio
(\ref{eq:a45}).

For $H \rightarrow b\bar{b}$, the signal-to-QCD background ratio
is
\begin{equation}
\label{eq:a46}
 \frac{S (gg^{PP} \rightarrow b\bar{b})}{B (gg^{PP} \rightarrow
 gg)}\; \simeq \; 4.3 \times 10^{-3} \: {\rm Br} (H \rightarrow
 b\bar{b}) \: \left ( \frac{M}{100~{\rm GeV}} \right )^3 \: \left (
 \frac{250~{\rm MeV}}{\Delta M} \right ),
\end{equation}
which is very small \cite{KMRmm}.  Fortunately, if we tag the $b$
and $\bar{b}$ jets to reject the $gg$ events, we can strongly
suppress the QCD background.  Recall that the $gg^{PP} \rightarrow
b\bar{b}$ QCD background process is suppressed by colour and spin
factors, and by the $J_z = 0$ selection rule.  In this way the
background is suppressed by an extra factor
\begin{equation}
\label{eq:a47}
 \frac{m_b^2}{M_H^2} \: \frac{1}{4} \: \frac{1}{27}
 \; \lapproxeq \; 2 \times 10^{-5}.
\end{equation}
However the full suppression is only true in the Born
approximation.  Recall that large angle gluon radiation in the
final state violates the $J_z = 0$ selection rule \cite{SUP}, so
in the $b\bar{b}$ QCD background the suppression factor
$m_b^2/M_H^2$ is replaced by $\alpha_S/\pi$.  The final result for
the $H \rightarrow b\bar{b}$ signal-to-background ratio is
therefore
\begin{equation}
\label{eq:a48}
 \frac{S (gg^{PP} \rightarrow H \rightarrow b\bar{b})}{B (gg^{PP} \rightarrow
 b\bar{b})} \; \gapproxeq \; 15 \left ( \frac{250~{\rm MeV}}{\Delta
 M} \right ),
\end{equation}
for a Higgs boson of mass $M_H = 120$~GeV.

\subsection{Double-diffractive $\gamma\gamma$ production}

At first sight, the subprocess $gg^{PP} \rightarrow \gamma\gamma$
appears attractive to serve as an alternative $gg^{PP}$ luminosity
monitor for the exclusive double-diffractive processes\footnote{We
are grateful to Mike Albrow and Beate Heinemann for discussions on
this proposal.}. However it turns out that the event rate is too
small. Using the known results\footnote{We thank Andrei Shuvaev
for checking various formulae for $\gamma\gamma$ production.} for
the QED $\gamma\gamma \rightarrow \gamma\gamma$ helicity
amplitudes \cite{GAMGAM}, we calculate the $J_z = 0$, parity $P =
+1$ subprocess cross section to be
\begin{equation}
\label{eq:a49}
 \hat{\sigma} (30^\circ < \theta_\gamma^* < 150^\circ) \; \simeq
 \; 0.3 (0.04)~{\rm pb}
\end{equation}
for $M_{\gamma\gamma} \sim 50 (120)$~GeV.  When multiplied by the
luminosity at the Tevatron of $1.8 \times 10^{-3} (2 \times
10^{-4})$, we obtain, assuming $\int dM^2/M^2 \simeq 1$,
\begin{equation}
\label{eq:b49}
 \sigma (p\bar{p} \rightarrow p + \gamma\gamma + \bar{p}) \;
 \simeq \; 0.5~{\rm fb} \; (0.008~{\rm fb}).
\end{equation}
On the other hand, for the LHC for $M_{\gamma\gamma} \simeq
120$~GeV we expect
\begin{equation}
\label{eq:c49}
 \sigma (pp \rightarrow p + \gamma\gamma + p) \; \simeq \;
 0.12~{\rm fb}.
\end{equation}
Note that for $M_{\gamma\gamma} \simeq 120$~GeV the observable
rate of Standard Model exclusive $\gamma\gamma$ events is expected
to be much less than even Higgs $\rightarrow b\bar{b}$ signal.  In
comparison to exclusive $gg^{PP} \rightarrow gg$ subprocess, the
$gg^{PP} \rightarrow \gamma\gamma$ rate is smaller by a factor of
$10^6$ for $E_{Tg} \simeq E_{T\gamma}$.

This obvious disadvantage of the Standard Model exclusive diphoton
production can be turned into an attractive advantage for searches
at hadronic colliders of signs of the existence of extra
dimensions, which for instance appear in theories of low-scale
gravity \cite{ADD}. The small background rate, induced by the
conventional box diagram contribution, may allow a very sensitive
high-mass diphoton probe of the effective scale of quantum gravity
effects, $M_S$ (see, for example, \cite{KC,HD}). To illustrate the
level of the possible signal we follow the estimates of
Ref.~\cite{KC} with effective scale $M_S = 1.5$~TeV. At
$M_{\gamma\gamma} \sim 500$~GeV one then would expect, for
$30^\circ < \theta_\gamma^* < 150^\circ$,
\begin{equation}
\label{eq:d49}
 \hat{\sigma}^{\rm incl}_{\rm extra~dim.} \; \sim \; F^2 \left ( \int \frac{dM^2}{M^2}
 \right ) \: 50~{\rm fb},
\end{equation}
where the factor $F$ is given by (see, for example, \cite{KC})
\begin{equation}
\label{eq:e49}
 F \; = \; 2/(n - 2) \quad {\rm for} \quad n \; > \; 2,
\end{equation}
where $n$ is the number of extra (compactified) dimensions.  We
consider here the inclusive configuration of Fig.~1(b), since for
exclusive LO kinematics there is no $J_z = 0$ point-like
gluon-gluon coupling to a $2^+$ ``graviton''.

Taking $\int dM^2/M^2 \sim 1$, after multiplying by the
double-diffractive inclusive luminosity of $2 \times 10^{-3}$ (for
$\Delta \eta = 3$), we would expect at the LHC
\begin{equation}
\label{eq:f49}
 \Delta \sigma_{\rm extra~dim.} (pp \rightarrow X + \gamma\gamma +
 Y) \; \sim \; 0.1 F^2~{\rm fb},
\end{equation}
which, assuming that $F^2$ is of order 1, should yield about 10
events, for an integrated luminosity of ${\cal L} = 100~{\rm
fb}^{-1}$. It is quite possible that future Tevatron results may
increase the existing lower limit on $M_S$, and hence make a
diphoton signature for extra dimensions invisible.

Of course, in the inclusive configuration, there may be some
additional background induced by $\gamma\gamma$ emission off the
quark lines in the process where the rapidity gaps are created by
colourless $q\bar{q}$ $t$-channel exchange.  However for gaps with
$\Delta \eta = 3$, a preliminary estimate\footnote{We plan to
study $q\bar{q}$ exchange in more detail in the future.} shows
that this quark contribution does not exceed the background
contribution coming from the $gg \rightarrow \gamma\gamma$ box
diagram.  Thus the total background should not be greater than
about 5--10\%.

\subsection{$t\bar{t}$ production}

The exclusive double-diffractive production of $t\bar{t}$ pairs
may provide new opportunities for studying top quark physics.  In
particular, it offers a novel probe of the QCD dynamics in the
$t\bar{t}$ system.  The Born cross section is given by
(\ref{eq:a40}). Due to parity conservation, the $t\bar{t}$ pair is
produced in a $P$-wave state, and so the threshold behaviour goes
as $\beta^3$, where $\beta$ is the quark velocity in the
$t\bar{t}$ rest frame. Of course, the Born cross section is
modified by Coulomb and top-quark width effects \cite{TOP,FK1},
but for illustration it is sufficient to neglect these, see our
estimates at the end of subsection 3.5.2.

Note that the observation of $t\bar{t}$ exclusive production can
be used as a template of the ability of the missing mass method to
find signals of New Physics.

\subsection{SUSY particle production}

Double-diffractive exclusive processes, in principle, provide a
unique opportunity to investigate the whole of the strong
interaction sector of physics beyond the Standard Model.  As an
example, we consider supersymmetry, which is a front-runner in
searches for New Physics.  Here we ignore the current theoretical
SUSY prejudices and assume for illustration, first, that the
gluino is the lightest supersymmetric particle (LSP).  The second
scenario that we consider is that unstable gluinos and squarks
exist with masses close to the current experimental limits.  Note
that it is difficult to separate gluinos and squarks produced in
standard inelastic hadronic collisions.  Moreover, gluino studies
appear especially promising in $gg$ collisions since these cannot
be easily achieved at linear $e^+ e^-$ colliders.

\subsubsection{Gluinoball production}

In recent publications \cite{MAFI} an interesting possibility is
discussed that a gluino $\tilde{g}$ is the LSP (or next-to-LSP)
or, to be specific, that it does not decay within the detector.
The allowed mass window is claimed to be 25--35~GeV. Within such a
scenario there should exist a spectrum of $\tilde{g}\tilde{g}$
bound states (gluinoballs or gluinonia), which may reveal
themselves in gluon-gluon collisions. The orbital angular momentum
$L$ must be odd in such $gg^{PP}$ collisions, due to parity
conservation. Therefore the lowest lying colourless bound states
is the $0^{++} (^3 P_0)$ state, which we denote $\tilde{G}$, with
principal quantum number $n = 2$. Relative to the $2m_{\tilde{g}}$
threshold, the energies of such $P$-wave $\tilde{g}\tilde{g}$
bound states are
\begin{equation}
\label{eq:a50}
 E_n \; = \; - \: \frac{9}{4} \: m_{\tilde{g}} \: \frac{\alpha_S^2}{n^2}
\end{equation}
with $n \geq 2$, where $\alpha_S$ is to be evaluated at the
Coulombic scale $k_G = 3 \alpha_S m_{\tilde{g}}/2$ (see, for
example, \cite{GH,BFK}). We may estimate the partial width of the
$\tilde{G} \rightarrow gg$ decay using the Coulombic
approximation.  The old results of the two-photon decay of
$P$-wave positronium \cite{POS} may be applied (see, for example,
\cite{GH,BFK}), which gives
\begin{eqnarray}
\label{eq:a51}
 \Gamma (\tilde{G} \rightarrow gg) & = & 6.4 \: \alpha_S^2
 (m_{\tilde{g}}) \: \alpha_S^5 (k_G) \: M_{\tilde{G}} \nonumber \\
 & & \\
 & \simeq & \left [ \frac{M_{\tilde{G}}}{60~{\rm GeV}}\right ] \: 0.2~{\rm
 MeV}, \nonumber
\end{eqnarray}
where $M_{\tilde{G}} \approx 2 m_{\tilde{g}}$.  Then, on using
(\ref{eq:a24}), we obtain
\begin{equation}
\label{eq:a52}
 \hat{\sigma}^{\rm excl} (gg^{PP} \rightarrow \tilde{G}) \; = \;
 \delta \left ( 1 \: - \: \frac{M^2}{M_{\tilde{G}}^2}\right ) \; 18~{\rm
 pb}.
\end{equation}
The anticipated cross section for the signal is quite sizeable.
For example, for $M_{\tilde{G}} = 60$~GeV,
\begin{eqnarray}
\label{eq:b52}
 \sigma (pp \rightarrow p + \tilde{G} + p) & \simeq & 0.4~{\rm pb}~({\rm
 LHC}), \nonumber \\
 & & \\
 \sigma (p\bar{p} \rightarrow p + \tilde{G} + \bar{p}) & \simeq &
 20~{\rm fb}~({\rm Tevatron}). \nonumber
\end{eqnarray}
However the signal-to-background ratio is
\begin{equation}
\label{eq:a53}
 \frac{S (gg^{PP} \rightarrow \tilde{G} \rightarrow gg)}{B (gg^{PP} \rightarrow
 gg)} \; = \; 0.6 \times 10^{-2} \left ( \frac{250~{\rm MeV}}{\Delta
 M}\right ) \left ( \frac{M_{\tilde{G}}}{60~{\rm GeV}} \right ),
\end{equation}
which makes detection difficult, even with angular cuts. Some
words of caution are in order here.  Our estimates are based
simply on the lowest-order Coulombic formula, (\ref{eq:a51}). More
refined calculations of $\Gamma (\tilde{G} \rightarrow gg)$ are
certainly needed.  Secondly, we have neglected the possible
interference between the signal and the background.

The threshold production of (quasi) stable gluino pairs may be
strongly affected by QCD final-state interactions, in analogy to
the celebrated Coulomb threshold phenomena in QED \cite{AS}.  As
follows from the results of Refs.~\cite{FK1,BFK}, in the
zero-width approximation the $P$-wave threshold cross section,
$d\sigma_P (\tilde{g}\tilde{g}; M)$, is
\begin{equation}
\label{eq:b53}
 d \sigma_P (\tilde{g}\tilde{g}; M) \; = \; d\sigma_P^{(0)}
 (\tilde{g}\tilde{g}; M) \: |\psi_{\tilde{g}} (0)|^2 \: \left (1
 \: + \: \frac{Z_g^2}{4\pi^2} \right ),
\end{equation}
where $d\sigma_P^{(0)} (\tilde{g}\tilde{g}, M)$ denotes the Born
cross section at cm energy $M \geq 2 m_{\tilde{g}}$.
$\psi_{\tilde{g}} (0)$ is the Coulombic wave function of a
colour-singlet $\tilde{g}\tilde{g}$ $S$-wave state evaluated at
the origin (c.f.\ Refs.~\cite{AS,AP}).
\begin{equation}
\label{eq:c53}
 |\psi_{\tilde{g}} (0)|^2 \; = \; \frac{Z_g}{1 - \exp (-Z_g)}
\end{equation}
with
\begin{equation}
\label{eq:d53}
 Z_g \; = \; \frac{3 \pi \alpha_S (p_C)}{\beta_{\tilde{g}}},
\end{equation}
where $\beta_{\tilde{g}} \; = \; \sqrt{1 - 4 m_{\tilde{g}}^2/M^2}$
is the velocity of the $\tilde{g}$.  It looks natural to choose
the scale of $\alpha_S$ in (\ref{eq:d53}) to be $p_C = {\rm max}
\left \{ \left (m_{\tilde{g}} (M - 2 m_{\tilde{g}}) \right
)^{1/2}, (m_{\tilde{g}} \Delta M)^{1/2} \right \}$.

Since $Z_g \gapproxeq 1$, QCD Coulombic effects drastically modify
the $\tilde{g}\tilde{g}$ excitation curve, strongly enhancing the
production rate at or near threshold.  According to
(\ref{eq:b53}), for $Z_g^2 < 4 \pi^2$ the threshold cross section
rises with increasing $\beta_{\tilde{g}}$ approximately as
$\beta_{\tilde{g}}^2$, rather than $\beta_{\tilde{g}}^3$ as one
would expect from the Born result.  Moreover, for $Z_g^2 \gg 4
\pi^2$, the colour Coulombic exchanges in the final state
completely compensate the kinematical $\beta_{\tilde{g}}^3$
threshold factor contained in the Born cross section
\cite{FK1,BFK}
\begin{equation}
\label{eq:e53}
 \beta_{\tilde{g}}^3 \: \frac{d \sigma_P (\tilde{g}\tilde{g}; M)}{d\sigma_P^{(0)}
 (\tilde{g}\tilde{g}; M)} \; \rightarrow \; \frac{27 \pi}{4} \:
 \alpha_S^3 (p_C) \; \simeq \; 0.14,
\end{equation}
for $p_C \simeq 8$~GeV which corresponds to $\Delta M \simeq
1$~GeV.  This is a dramatic effect relative to the lowest-order
expectation.  The (LSP) gluino-pair production process would lead
to distinctive signatures in hadronic collisions, in particular to
events with jets accompanied by missing $p_T$, see, for details,
Ref.~\cite{MAFI,BCG}.

Despite all the attractive features of the gluino-LSP scenario
\cite{MAFI}, it is possible that direct experimental studies at
the Tevatron will close this window.

\subsubsection{Gluino and squark production}

A more plausible scenario is to search for massive unstable
gluinos $\tilde{g}$ or squarks $\tilde{q}$, which will reveal
themselves as multijet final states with missing energy and/or
leptons\footnote{At the moment we have no clear understanding
which of these sparticles is the lightest, but the conventional
belief is that the stop $\tilde{t}$ is the lightest squark.}. As
noted above the advantage of exclusive double-diffractive studies
is that gluinos and squarks can be separated by their respective
$\beta^3$ and $\beta$ threshold behaviour.  The Born subprocess
cross sections are
\begin{eqnarray}
\label{eq:a54}
 \frac{d\hat{\sigma}^{\rm excl}}{dt} (gg^{PP} \rightarrow
 \tilde{g}\tilde{g}) & = & \frac{27}{2} \: \frac{1}{2} \:
 \frac{\pi \alpha_S^2}{6 E_T^4} \: \frac{m_{\tilde{g}}^2}{M^2}
 \: \beta_{\tilde{g}}^2, \\
 & & \nonumber \\
 \label{eq:b54}
 \frac{d\hat{\sigma}^{\rm incl}}{dt} \: (gg^{PP} \rightarrow
 \tilde{g}\tilde{g}) & = & \frac{27}{2} \: \frac{1}{2} \:
 \frac{\pi \alpha_S^2}{6E_T^2 M^2} \left [ \left (1 -
 \frac{2E_T^2}{M^2}\right ) \left (1 -
 \frac{2m_{\tilde{g}}^2}{E_T^2}\right ) \: + \:
 \frac{m_{\tilde{g}}^2}{E_T^2}\left (1 + \beta_{\tilde{g}}^2
 \right ) \right ], \\
 & & \nonumber \\
 \label{eq:a55}
 \frac{d \hat{\sigma}^{\rm excl}}{dt} \: (gg^{PP} \rightarrow
 \tilde{q} \tilde{\bar q}) & = & \frac{4 \pi \alpha_S^2}{12 M^4} \:
 \frac{m_{\tilde{q}}^4}{E_T^4}, \\
 & & \nonumber \\
 \label{eq:b55}
 \frac{d\hat{\sigma}^{\rm incl}}{dt} \: (gg^{PP} \rightarrow
 \tilde{q}\tilde{\bar q}) & = & \frac{2 \pi \alpha_S^2}{12 M^4}
 \left ( 1 \: - \: \frac{2m_{\tilde{q}}^2}{E_T^2} \: + \:
 \frac{2 m_{\tilde{q}}^4}{E_T^4} \right ).
\end{eqnarray}
Note that (\ref{eq:a54}) and (\ref{eq:b54}) are the same as
(\ref{eq:a40}) and (\ref{eq:a42}) for colour-singlet $gg
\rightarrow q\bar{q}$, except for the colour-factor $27/2$, and a
factor $1/2$ which reflects the Majorana nature of the two
gluinos.  The $\beta^3$ behaviour of exclusive gluino-pair
production arises from $\beta^2$ of the last factor of
(\ref{eq:a54}) and $\beta$ from the integration over $t$.  The
cross sections shown in (\ref{eq:a55}) and (\ref{eq:b55}) are
written for one flavour and one type ($L$ or $R$) of squark
$\tilde{q}$.

Near threshold the formula have to be modified to account for
sparticle widths and the QCD Coulomb interaction \cite{TOP,FK1}.
Unfortunately the cross sections are very small near threshold.
For example, if we take sparticle masses of 250~GeV and integrate
from threshold (500~GeV) up to 625~GeV, then we find
\begin{equation}
\label{eq:a56}
 \Delta \hat{\sigma}^{\rm excl} (\tilde{g}\tilde{g}) \; \simeq \;
 6.5~{\rm pb}, \quad\quad\quad \Delta \hat{\sigma}^{\rm excl}
 (\tilde{q}\tilde{\bar q}) \; \simeq \; 1.8~{\rm pb}.
\end{equation}
For comparison the $t\bar{t}$ cross section, integrated from
threshold up to 437~GeV, corresponding to the same value of $\beta
= 0.6$, is
\begin{equation}
\label{eq:a57}
 \Delta \hat{\sigma}^{\rm excl} (t\bar{t}) \; \simeq \; 2.2~{\rm
 pb}.
\end{equation}
After multiplying by the double-diffractive luminosities of $2
\times 10^{-5}$ for $\tilde{g}\tilde{g}$ and $4 \times 10^{-5}$
for $t\bar{t}$ we have
\begin{eqnarray}
\label{eq:a58}
 \Delta \sigma (pp \rightarrow p + \tilde{g}\tilde{g} + p) &
 \simeq & 0.15~{\rm fb} \\
 & & \nonumber \\
 \Delta \sigma (pp \rightarrow p + \tilde{q}\tilde{\bar q} + p) &
 \simeq & 0.04~{\rm fb} \\
 & & \nonumber \\
 \label{eq:a59}
 \Delta \sigma (pp \rightarrow p + t\bar{t} + p) & \simeq &
 0.1~{\rm fb}
\end{eqnarray}
for `near' threshold production at the LHC.  It is quite plausible
that the masses of the light squark flavours are nearly
degenerate, as are the masses of $\tilde{q}_L$ and $\tilde{q}_R$.
This may allow a higher rate of $\tilde{q}\tilde{\bar q}$ events.

Another possibility to increase the yield of SUSY production is to
consider the inclusive configuration of Fig.~1(b).  Then the
effective luminosity $(\sim 2 \times 10^{-3})$ is much larger. The
corresponding subprocess cross sections, integrated from threshold
(500~GeV) up to 625~GeV, are
\begin{equation}
\label{eq:b59}
 \Delta \hat{\sigma}^{\rm incl} (\tilde{g}\tilde{g}) \; \simeq \;
 24~{\rm pb}, \quad\quad\quad \Delta \hat{\sigma}^{\rm incl}
 (\tilde{q}\tilde{\bar q}) \; \simeq \; 1~{\rm pb},
\end{equation}
leading to
\begin{eqnarray}
\label{eq:c59}
 \Delta \sigma (pp \: \rightarrow \: X \: + \: \tilde{g}\tilde{g}
 \: + \: Y) & \simeq & 50~{\rm fb}, \\
 & & \nonumber \\
\label{eq:d59}
 \Delta \sigma (pp \: \rightarrow \: X \: + \:
 \tilde{q}\tilde{\bar q} \: + \: Y) & \simeq & 2~{\rm fb}.
\end{eqnarray}

\subsection{Soft Phenomena}

Finally, we discuss the possibility to study soft strong
interactions in diffractive events with two (or more) rapidity
gaps (see, for example, \cite{ALRK,KAID}).  The (total) cross
section of the Pomeron-Pomeron interaction is proportional to the
square of the triple Pomeron coupling $g_{3\funp}$.  If we take
the parameters from the analysis of Ref.~\cite{KMRsoft} then we
obtain the subprocess cross section
\begin{equation}
\label{eq:a60}
 \hat{\sigma}_t (\funp\funp) \; \simeq \; 1.2~{\rm mb}.
\end{equation}
When we multiply by the ``soft $\funp\funp$'' luminosity of
Section~2.3, shown in Fig.~2, we predict a two rapidity gap cross
section
\begin{equation}
\label{eq:a61}
 \sigma_2 \; \sim \; 1-10~\mu{\rm b},
\end{equation}
depending on the gap size.  For the LHC it gives a very large
event rate, which may be measured in a low luminosity (high
$\beta$ optics) run.

Such events offer an excellent opportunity to answer the many
outstanding, interesting questions which remain from the intensive
studies of soft physics of 25 or more years ago (see, for example,
\cite{KAID,AKLR}). A key problem is that the cross section of
events containing $n$ rapidity gaps is of the form
\begin{equation}
\label{eq:a62}
 \sigma_n \; \sim \; (g_{3\funp} \ln s)^{2n}/(2n)!,
\end{equation}
which grows with energy.  On summing, we have a cross section
$\sigma \sim (s)^{g_{3\funp}}$, which appears to violate
unitarity.  We must find a mechanism to suppress this huge cross
section.  We recall the scenarios which were considered long ago.
\begin{itemize}
\item[(i)] To obtain a self consistent asymptotic theory it was
proposed that the triple Pomeron coupling vanishes when the
momentum transferred through the Pomeron goes to zero,
\begin{equation}
\label{eq:a63}
 g_{3\funp} (t) \; \rightarrow \; 0 \quad\quad {\rm as} \quad\quad
 t \; \rightarrow \; 0.
\end{equation}
The diffractive data do not confirm this proposal, although in the
presence of strong $\funp$-cuts it is still an open question.  The
study of double-Pomeron-exchange processes in low luminosity LHC
runs, where the small $t$ region becomes accessible, would clarify
the situation (see \cite{AKLR} and references therein).

\item[(ii)] It was suggested that the soft survival probability
$\hat{S}^2$ becomes much smaller for events with many gaps.  The
available CDF data \cite{CDF}, on double and single diffractive
production, indicate otherwise, but again it should be checked at
LHC energies.

\item[(iii)] It is more natural to expect $\hat{S}^2$ does not
depend strongly on the number of gaps, but rather to propose that
$\hat{S}^2$ decreases rapidly with increasing energy.  In this
scenario, rapidity gap events {\it only} occur in peripheral
(large impact parameter) collisions where the opacity $\Omega (b)$
is still small.  It follows that mean transverse momentum of the
secondaries $\langle k_t \rangle$ created in the central
$\funp\funp$ collisions of Fig.~1(c) will be smaller than $\langle
k_t \rangle$ for the usual inelastic interaction at an equivalent
energy, $\sqrt{s_{\rm inel}} = M$ \cite{RYSKIN}.  This prediction
could be checked at the LHC.
\end{itemize}

We emphasize that processes with 3 or more rapidity gaps have not
yet been seen.  It is clearly crucial to detect such events.  To
estimate $\sigma_3, \sigma_4, \ldots$, we assume factorization,
and the same survival factor $\hat{S}^2$ for $\sigma_1, \sigma_2,
\sigma_3, \ldots$.  Then the probability to observe an additional
rapidity gap will be
\begin{equation}
\label{eq:a64}
 \omega_{\rm gap} \; = \; \left ( \frac{\sigma^{DD}}{\sigma_{\rm inel}
 \hat{S}^2} \right ),
\end{equation}
where we cancel the factor $\hat{S}^2$ in the empirical
double-diffractive cross section $\sigma_{DD}$, since the survival
factor $\hat{S}^2$ is the same for all $\sigma_n$.  Of course, we
have to account for the phase space available for each gap.  To
make a phenomenological estimate, we start with the two-gap
process of Fig.~1(c) and evaluate the probability to observe a
third gap at the Pomeron-Pomeron energy $\sqrt{s_{\funp\funp}} =
M$.  For the LHC it corresponds to $\sqrt{s_{\funp\funp}} \sim
300$~GeV.  Using the existing data in the ISR to Tevatron energy
range, together with the $\hat{S}^2$ calculations of
Ref.~\cite{KMRsoft}, we find
\begin{equation}
\label{eq:a65}
 \omega_{\rm gap} \; \simeq \; 0.2-0.4
\end{equation}
and
\begin{equation}
\label{eq:a66}
 \sigma_3 \; \simeq \; \omega_{\rm gap} \sigma_2 \; \sim \;
 0.3-3~\mu{\rm b}.
\end{equation}

Another possible soft phenomenon which is worth investigation is
the ``elastic'' Pomeron-Pomeron scattering process
\begin{equation}
\label{eq:a67}
 \funp\funp \; \rightarrow \; X^G Y^G,
\end{equation}
where $X^G$ and $Y^G$ are glueball states.  We call it elastic as
the glueballs may lie on the Pomeron trajectory.  That is there
may exist a glueball dominance model of the Pomeron, in analogy to
the vector meson dominance model of the photon.

\section{Conclusions}

In this paper we summarize, and extend, the results of our
previous studies of double-diffractive processes with a rapidity
gap on either side of a centrally produced heavy system.  We have
attempted to present the results in a unified, and user-friendly,
form so that the expected cross sections at the Tevatron and the
LHC can be readily estimated, simply by multiplying the
appropriate luminosity of Section~2 with the relevant subprocess
cross section $\hat{\sigma}$ of Section~3.  We hope that it
provides a useful framework to study double-diffractive phenomena,
not only in their own right but also as a tool to probe physics
beyond the Standard Model.  Although many cross sections are
predicted to be small, there are several interesting processes
with viable signals.  In particular, the exclusive
double-diffractive production of an intermediate mass Higgs boson
at the LHC, and the possibility of using the Tevatron and the LHC
as a `pure' gluon factory. For example, for the production of a
Higgs boson of mass $M_H = 120$~GeV at the LHC we predict a cross
section
\begin{equation}
\label{eq:a68}
 \sigma (pp \rightarrow p + H + p) \; \simeq \; 3~{\rm fb},
\end{equation}
with a signal-to-QCD background ratio of
\begin{equation}
\label{eq:a69}
 \frac{S (H \rightarrow b\bar{b})}{B (b\bar{b})} \; \gapproxeq \;
 15 \left ( \frac{250~{\rm MeV}}{\Delta M} \right ),
\end{equation}
where $\Delta M$ is the experimental missing-mass resolution.

\section*{Acknowledgements}

We thank Mike Albrow, Brian Cox, Gian Giudice, Dino Goulianos,
Beate Heinemann, Aliosha Kaidalov, Risto Orava, Robi Peschanski,
Krzysztof Piotrzkowski, Stuart Raby, Albert de Roeck, Andrei
Shuvaev, Stefan Tapprogge and Georg Weiglein for interesting
discussions. One of us (VAK) thanks the Leverhulme Trust for a
Fellowship. This work was partially supported by the UK Particle
Physics and Astronomy Research Council, by the Russian Fund for
Fundamental Research (grants 01-02-17095 and 00-15-96610) and by
the EU Framework TMR programme, contract FMRX-CT98-0194 (DG
12-MIHT).

\end{document}